\documentclass[epj,nopacs,final]{svjour}

 \usepackage{latexsym}
 \usepackage{graphics}
 \usepackage{epsfig}
 \usepackage{amsmath}
 \usepackage{amssymb}
 \usepackage{array}
 \usepackage[dvips]{color}
 \usepackage{afterpage}
 \usepackage{url}
 
\begin{document}

\title{Quasielastic axial-vector mass from experiments on neutrino--nucleus scattering.}

\author{K.~S.~Kuzmin\inst{1,2,}\thanks{\email{KKuzmin@jinr.theor.ru}}                    \and
        V.~V.~Lyubushkin\inst{3,4,}\thanks{\email{Vladimir.Lyubushkin@cern.ch}}          \and
        V.~A.~Naumov\inst{1,}\thanks{\email{VNaumov@jinr.theor.ru}}
        }
\institute{Bogoliubov Laboratory of Theoretical Physics,
           Joint Institute for Nuclear Research, RU-141980 Dubna, Russia                 \and
           Institute for Theoretical and Experimental Physics, RU-117259 Moscow, Russia  \and
           Dzhelepov Laboratory of Nuclear Problems,
           Joint Institute for Nuclear Research, RU-141980 Dubna, Russia                 \and
           Physics Department of Irkutsk State University, RU-664003, Irkutsk, Russia
          }

 \date{Received: December 29, 2007}


\abstract{
We analyse available experimental data on the total and differential
charged-current cross sections for quasielastic $\nu_{\mu}N$ and
$\overline\nu_{\mu}N$ scattering, obtained with a variety of nuclear targets
in the accelerator experiments at ANL, BNL, FNAL, CERN, and IHEP, dating from
the end of sixties to the present day. The data are used to adjust the poorly
known value of the axial-vector mass of the nucleon.
}

\maketitle

\section{Introduction}

A precise knowledge of the cross sections for charged-current induced
quasielastic scattering (QES) of neutrinos and antineutrinos on nuclear targets
is a pressing demand of the current and planning next generation experiments with
accelerator and atmospheric neutrino beams, aiming at the further exploration
of neutrino oscillations, probing nonstandard neutrino interactions, searches for
proton decay, and related phenomena. 

The quasielastic cross sections are very sensitive to the poorly known shape of the
weak axial-vector form factor $F_A(Q^2)$ of the nucleon.
Adopting the conventional dipole approximation, this form factor is determined
by the axial-vector coupling $g_A=F_A(0)$ and the phenomenological parameter $M_A$,
the so-called axial-vector (dipole) mass related to the root-mean-square axial radius by
\[
\langle{r_A^2}\rangle=-\frac{6}{g_A}\left[\frac{dF_A(Q^2)}{dQ^2}\right]_{Q^2=0}
  =\frac{12}{M_A^2}.
\]
The experimental values of $M_A$ extracted from neutrino and antineutrino scattering
data and from the more involved and vastly model-dependent analyses of charged pion
electroproduction off protons, show very wide spread, from roughly 0.7 to 1.2~GeV
with the formal weighted averages \cite{Liesenfeld:99,Bernard:01} 
\[
M_A=\left\{
\begin{aligned}
1.026 & \pm 0.021~\text{GeV} \enskip \text{from $\nu_\mu,\overline\nu_\mu$ experiments}, \\
1.069 & \pm 0.016~\text{GeV} \enskip \text{from $\pi$ electroproduction}.
\end{aligned}
\right.
\]
The first value, the common default in most current neutrino simulations, is defined largely by
$\nu_{\mu}d$ bubble chamber experiments; in many of these experiments, the extractions of $M_A$
were based on the naive dipole approximation for the vector form factors of the nucleon,
along with other conjectures. 
The second value should be in fact decreased by about 5\%, in order to account for hadronic loop
corrections (see, e.g., Ref.\ \cite{Liesenfeld:99}).

The results of several selected $\nu_{\mu}d$, $\overline\nu_{\mu}\text{H}$, and $\pi^\pm$
electroproduction experiments have been recently reanalyzed by Bodek et al.\ \cite{Bodek:07c},
using a new improved description of the vector form factors (``BBBA(07)'' parametrization).
The obtained world average axial mass is
\[
M_A=1.014 \pm 0.014~\text{GeV} \quad \text{(BBBA(07))}.
\]
This value seems to be in conflict with the new results of high-statistics neutrino experiments
K2K SciFi \cite{Gran:06} (oxygen target) and MiniBooNE \cite{Aguilar-Arevalo:07} (carbon target),
reported unexpectedly large while mutually consistent values of the axial mass:
\[
M_A=\left\{
\begin{aligned}
1.20 & \pm 0.12~\text{GeV} \quad \text{(K2K)}, \\
1.23 & \pm 0.20~\text{GeV} \quad \text{(MiniBooNE)}
\end{aligned}
\right.
\]
A preliminary analysis of antineutrino data in MiniBooNE yields a consistent value of $M_A$ \cite{Katori:07}.

Both K2K and MiniBooNE extractions utilize the \emph{updated} vector form factors, from Refs.\
\cite{Bosted:95,Budd:03} and \cite{Bradford:06}, respectively.
Within the low-$Q^2$ regions explored in K2K and MiniBooNE experiments, the difference between these
parametrizations and BBBA(07) is comparatively small.

It can be noted that nuclear effects in the K2K analysis were accounted within the standard relativistic
Fermi gas (RFG) model \cite{Smith:72}, while the MiniBooNE analysis used RFG modified by including an
``instrumental'' free parameter $\kappa$ which changes the strength of Pauli-blocking.
A fit of the $Q^2$ shape above $0.25~\text{GeV}^2$ (where the variations of $\kappa$ has no significant impact)
leads to an even larger value of $M_A=1.25\pm0.12~\text{GeV}$.

In this study, which is in a sense complementary to that by Bodek et al.\ \cite{Bodek:07c}, we attempt
to extract the axial mass value by a global statistical analysis of all available consistent data on the
total and differential QES cross sections measured in accelerator experiments
with $\nu_\mu$ and $\overline\nu_\mu$ beams%
\footnote{The $\nu_e$, $\overline\nu_e$, $\nu_\tau$, and $\overline\nu_\tau$ beams from past and current
          accelerator experiments are not appropriate for measuring the QES cross sections.}
from
ANL  \cite{Novey:67,Kustom:69a,Mann:72,Mann:73,S.Barish:75,Singer:77,%
           S.Barish:77a,Miller:82},
BNL  \cite{Cazzoli:76,Cnops:78a,Fanourakis:80,Baker:81b,Kitagaki:86a,%
           Kitagaki:86b,Abe:86,Ahrens:87,Ahrens:88,Kitagaki:90,Sakuda:03,Furuno:03},
FNAL \cite{Asratyan:82,Kitagaki:83,Asratyan:84a,Asratyan:84b,Ammosov:86,Ammosov:87b,Suwonjandee:04},
CERN \cite{Burmeister:65,Franzinetti:66,Young:67,Orkin-Lecourtois:67,%
           Holder:68,Budagov:69c,Eichten:72,Eichten:73a,Sciulli:74a,%
           Haguenauer:74,Rollier:75,Deden:75,Bonetti:77,Erriquez:77,%
           Rollier:78,Dewit:78,Erriquez:78a,Pohl:79b,Armenise:79a,%
           Allasia:90,Petti:04,Lyubushkin:06,Martinez:07,Lyubushkin:08a}, and
IHEP \cite{Makeev:81,Belikov:81,Belikov:82a,Belikov:82b,Belikov:83,%
           Belikov:85a,Belikov:85b,Grabosch:86b,Grabosch:88,Brunner:90}.
The detector media used in these experiments are hydrogen, deuterium, carbon, aluminium, argon, iron/steel,
propane, freon, and also propane--freon and neon--hydrogen mixtures. 

In the likelihood analysis, we use the most accurate phenomenological parametrizations for the vector form factors
of the nucleon \cite{Bodek:07ab,Lomon:06}, we take into account all known sources of uncertainties, in particular,
the systematic errors in the energy spectra of ${\nu}_\mu$ and $\overline{\nu}_\mu$ beams. For description
of nuclear effects we apply the standard RFG model.
We examine possible difference between the values of $M_A$ extracted from $\nu_\mu$ and $\overline\nu_\mu$ data,
and cross-check our results with the data on $Q^2$ distributions measured in several experiments.

\section{Quasielastic neutrino scattering off free nucleon}
\label{sec:SM_model}


\subsection{Structure functions and cross section}

Let us first summarize the well-known phenomenology for describing the hypercharge conserved
quasielastic reactions on free nucleon targets
\begin{equation}\label{free}
\begin{gathered}
         \nu_{\ell}(k)+n(p)\to\ell^{-}(k')+p(p'), \\
\overline\nu_{\ell}(k)+p(p)\to\ell^{+}(k')+n(p').
\end{gathered}
\end{equation}
Here $k$, $k'$, $p$, and $p'$ denote the four-momenta and $\ell$ stays for $e$, $\mu$, or $\tau$.
In this paper, we will neglect the proton-neutron mass difference,%
\footnote{While our computer code operates with the most general formulas and relevant kinematics.}
since the resulting correction, in the $\nu_\mu/\overline\nu_\mu$ case, exclusively works near the
reaction threshold and practically negligible for the energies of our current interest.
The general formulas which take this effect into account, were derived in Ref.\ \cite{Strumia:03}
(assuming $T$ and $C$ invariance) and in Refs.\ \cite{Kuzmin:05b,Kuzmin:06b} (avoiding these assumptions).

The double differential cross-section for these processes is a convolution of
spin-averaged leptonic and hadronic tensors $L_{\alpha\beta}$ and $W_{\alpha\beta}$:
\begin{equation}\label{cs}
\frac{d\sigma_{\text{free}}}{dE_{\ell}d\cos\theta_\ell} = 
\frac{G_F^2P_\ell}{\pi(1+Q^2/M_W^2)^2}
\left(\frac{L^{\alpha\beta}W_{\alpha\beta}}{4ME_\nu}\right).
\end{equation} 
Here $G_F$ is the Fermi coupling,
$q=k-k'$ is the four-momentum transferred from the incoming (anti)neutrino to the nucleon,
$Q^2=-q^2$, $M_W$ is the mass of intermediate $W$-boson;
$E_\nu$, $E_\ell$, $P_\ell=\sqrt{E_\ell^2-m_\ell^2}$, and $\theta_\ell$ are, respectively,
the incident (anti)neutrino energy, outgoing lepton energy, momentum, and scattering angle
in the lab frame, $m_\ell$ is the lepton mass.
The leptonic tensor defined by the product of the weak leptonic currents, is given by
\begin{equation}\label{L}
L_{\alpha\beta}(k,k')=2\left[k'_{\alpha}k_{\beta}+k_{\alpha}k'_{\beta}-g_{\alpha\beta}(kk')
\mp i\varepsilon_{\alpha\beta\gamma\delta}k^{\gamma}k'^{\delta}\right],
\end{equation}
where the upper (lower) sign is for $\nu_\ell$ ($\overline\nu_\ell$).
Assuming the isotopic invariance, the hadronic tensor is defined by
the six structure functions $W_i(Q^2)$:
\begin{align} \label{W}
W_{\alpha\beta}(p,q) 
= & -g_{\alpha\beta} W_1+\frac{p_\alpha p_\beta}{M^2} W_2  \nonumber \\
  &   - \frac{i\varepsilon_{\alpha\beta\gamma\delta}p^\gamma q^\delta}{2M^2} W_3
      +  \frac{q_\alpha q_\beta}{M^2}W_4                   \nonumber \\
  &   +  \frac{p_\alpha q_\beta+q_\alpha p_\beta}{2M^2} W_5
      + i\frac{p_\alpha q_\beta-q_\alpha p_\beta}{2M^2} W_6,
\end{align}
where $M$ is the mass of the ``isoscalar'' nucleon.
Then combining Eqs.~\eqref{L} and \eqref{W} yields
\begin{align}\label{R}
\frac{L^{\alpha\beta}W_{\alpha\beta}}{4ME_\nu}
= & \left(\frac{E_\ell-P_\ell\cos\theta_\ell}{M}\right)
    \left(W_1+2\varkappa^2W_4\right)  \nonumber \\
&   \pm\left[\left(\frac{E_\nu+E_\ell}{M}\right) 
       \left(\frac{E_\ell-P_\ell\cos\theta_\ell}{2M}\right)
     -2\varkappa^2\right]W_3           \nonumber \\
&   +\frac{E_\ell+P_\ell\cos\theta_\ell}{2M}W_2-2\varkappa^2W_5,
\end{align}
where $\varkappa=m_\ell/2M$.

In order to connect the structure functions with the nucleon form factors,
we define the charged hadronic current for the QES process
(see, e.g., Ref.\ \cite{LlewellynSmith:72}):
\begin{equation}
\langle p(p')|J_\alpha|n(p)\rangle=V_{ud}\overline{u}_p(p')\Gamma_\alpha(p,q)u_n(p).
\end{equation}
Here $V_{ud}$ is the $ud$ transition element from the Cabibbo-Kobayashi-Maskawa
quark-mixing matrix and
\begin{align}
\Gamma_\alpha(p,q) = & \gamma_\alpha F_V
+i\sigma_{\alpha\beta}\frac{q^\beta}{2M}F_M+\frac{q_\alpha}{M}F_S \nonumber \\
                     & +\left(\gamma_\alpha F_A+\frac{p_\alpha+p'_\alpha}{M}F_T
                       +\frac{q_\alpha}{M}F_P \right)\gamma_5.
\end{align}
The form factors $F_i$ are in general complex functions of $Q^2$.
After standard calculations one finds
\begin{equation} \label{w_i}
W_i(Q^2)=2M^2|V_{ud}|^2w_i(Q^2)\delta\left(2(pq)-Q^2\right),
\end{equation}
with
\begin{align*}
w_1 = & |F_A|^2+x'\left(|F_V+F_M|^2+|F_A|^2\right),                            \\
w_2 = & |F_V|^2+|F_A|^2+x'\left(|F_M|^2+4|F_T|^2\right),                       \\
w_3 = & -2\mathrm{Re}\left[F_A^*(F_V+F_M)\right],                              \\
w_4 = & \frac{1}{4}\left[x'\left(|F_M-2F_S|+4|F_P+F_T|^2\right)-|F_M|^2\right] \\
      & +|F_S|^2+\frac{1}{2}\mathrm{Re}\left[F_V^*\left(2F_S-F_M\right)
                                                      -2F_A^*(F_P+F_T)\right], \\
w_5 = & w_2+2\mathrm{Re}\left[F_S^*\left(F_V-x'F_M\right)
                                         -F_T^*\left(F_A-2x'F_P\right)\right], \\
w_6 = & 2\mathrm{Im}\left[F_S^*\left(F_V-x'F_M\right)
                                         +F_T^*\left(F_A-2x'F_P\right)\right],
\end{align*}
and $x'=Q^2/4M^2$.
The only difference between this result and that from Ref.\ \cite{LlewellynSmith:72}
is in the relative sign of the terms in $\omega_6$ which does not contribute to the
QES cross section.%
\footnote{According to Llewellyn Smith, the functions $\omega_5'=\omega_5-\omega_2$
          and $\omega_6$ are, respectively, the real and imaginary parts of a unique
          function. Our examination does not confirm this property for the general
          case of nonvanishing second-class current induced form factors $F_S$ and $F_T$.
         }

Inserting Eqs.~\eqref{R} and \eqref{w_i} into Eq.~\eqref{cs} gives the commonly known
formula for the differential cross section for reactions \eqref{free} on free nucleon targets:
\begin{align*}
\frac{d\sigma_{\text{free}}}{dQ^2}
= & \frac{G_F^2M^2|V_{ud}|^2}{8\pi(1+Q^2/M_W^2)^2E_\nu^2} \nonumber \\
  & \times\left[A\frac{m_\ell^2+Q^2}{M^2}+B\frac{s-u}{M^2}
     +C\frac{(s-u)^2}{M^4}\right],
\end{align*}
where
\begin{align*}
A = & 2x'|F_V+F_M|^2-(1+x')|F_V|^2-x'(1+x')|F_M|^2                  \\
    &   +(1+x')|F_A|^2-4x'(1+x')|F_T|^2                             \\
    &   -\varkappa^2\left[|F_V+F_M|^2+|F_A+2F_P|^2 \right.          \\
    &    \left. -4(1+x')(|F_A|^2+|F_P|^2)\right],                   \\
B = & \mp 4x'\mathrm{Re}\left[F_A^*(F_V+F_M)\right]                 \\
    &   +4\varkappa^2\mathrm{Re}\left[F_T^*\left(F_A-x'F_P\right)
                               -F_S^*\left(F_V-x'F_M\right)\right], \\
C = & \frac{1}{4}\left(|F_V|^2+x'|F_M|^2+|F_A|^2+4x'|F_T|^2\right), \\
s = & \left(k +p\right)^2 = 2ME_\nu+M^2,                            \\
u = & \left(k'-p\right)^2 = m_\ell^2-2ME_\ell = m_\ell^2-2ME_\nu+Q^2.
\end{align*}

\subsection{Induced scalar and tensor form factors}

The quoted formulas take into account the nonstandard $G$ parity violating axial and vector
second-class currents (SCC)
which induce the nonzero scalar and tensor form factors $F_S$ and $F_T$. 
The most robust restrictions on the SCC couplings $F_{S,T}(0)$ come from the studies of
$\beta$ decay of complex nuclei (see, e.g., Refs.\ \cite{Wilkinson:00-01,Gardner:01} and
quoted therein references).
However, these studies are almost insensitive to the SCC effects at nonzero $Q^2$.
The latter were investigated in several (anti)neutrino experiments at BNL
\cite{Baker:81b,Abe:86,Ahrens:87,Ahrens:88} ($Q^2\lesssim1.2~\text{GeV}^2$) and
in the IHEP-ITEP spark chamber experiment at Serpukhov \cite{Belikov:85b}
($Q^2\lesssim2.4~\text{GeV}^2$), adopting the \emph{ad hoc} dipole parameterizations
\begin{align*}
F_S\left(Q^2\right)&=\xi_SF_V(0)\left(1+Q^2/M^2_S\right)^{-2}, \\
F_T\left(Q^2\right)&=\xi_TF_A(0)\left(1+Q^2/M^2_T\right)^{-2}.
\end{align*}
The strongest (but yet not too telling) 90\% C.L.\ upper limit for the axial SCC strength $\xi_T$
has been obtained at the BNL AGS $\overline\nu_\mu$ experiment \cite{Ahrens:88} as a function of
the ``tensor mass'' $M_T$, assuming conservation of vector current (CVC) (that is $\xi_S=0$),
and simple dipole form for the vector and axial form factors with $M_V=0.84~\text{GeV}$
and $M_A=1.09~\text{GeV}$. The limit ranges between 0.78 at $M_T=0.5~\text{GeV}$
to about 0.11 at $M_T=1.5~\text{GeV}$.
In so much as the contribution of the scalar form factor into the QES cross section is
suppressed by $(m_\mu/M)^2\approx0.01$, the 90\% C.L.\ constraint to the vector SCC strength
$\xi_S$ is even less impressive: $\xi_S<1.9$, assuming $\xi_T=0$, $M_S=1~\text{GeV}$, and the
same $M_V$ and $M_A$ as above.

Below, keeping in mind this vagueness, we will assume the time and charge invariance
of the hadronic current. Under this standard assumption, all the form factors are real
functions of $Q^2$ and
\[
F_S=F_T=0.
\]

\subsection{Vector form factors}

The Dirac and Pauli form factors $F_{V,M}$ are related to the Sachs electric and magnetic
form factors $G_{E,M}$:
\[
F_V=\frac{G_E+x'G_M}{1+x'},
\quad
F_M=\frac{G_M-G_E}{1+x'}.
\]
Isotopic symmetry provides simple relation between $G_{E,M}$ and elastic electric and magnetic
form factors of proton and neutron $G_E^{p,n}$ and $G_M^{p,n}$:
\[
G_M=G_M^p-G_M^n,
\quad
G_E=G_E^p-G_E^n.
\]
At low $Q^2$, a reasonable description of the electric and magnetic form factors is given by
the dipole approximation:
\begin{equation*}
G_E^p \approx       G_D,
\quad 
G_M^p \approx \mu_p G_D, 
\quad 
G_E^n \approx 0,  
\quad 
G_M^n \approx \mu_n G_D,
\end{equation*}
where $G_D=(1+Q^2/M_V^2)^{-2}$, $M_V=0.84~\text{GeV}$, and $\mu_p$ ($\mu_n$)
is the anomalous magnetic moment of the proton (neutron). Analyses of the almost all earlier neutrino
experiments were based on this approximation.
In this study, we utilize two more sophisticated models for the form factors $G_E^{p,n}$ and $G_M^{p,n}$
-- BBBA(07) \cite{Bodek:07ab} and GKex(05)\cite{Lomon:06}.

The BBBA(07) model is an accurate Kelly type parametrization of the current experimental data on the
form factors $G_E^p$, $G_M^p$, $G_E^n$, $G_M^n$, and ratio $G_E^p/G_M^p$, which uses the Nachtmann
scaling variable
\[
\xi_{p,n}=2\left(1+\sqrt{1+4M_{p,n}^2/Q^2}\right)^{-1}
\]
to relate elastic and inelastic form factors, and imposes quark-hadron duality asymptotic constraints
at high momentum transfers where the quark structure dominates.
The parametrization is based on the same datasets as were used by Kelly \cite{Kelly:04}, updated
to include some recent experimental results.
Quark-hadron duality implies that the squared ratio of neutron and proton magnetic form factors
should be the same as the ratio of the corresponding inelastic structure functions $F_2^n$ and
$F_2^p$ in the limit $\xi_{p,n}=1$:
\[
\left(\frac{G_M^n}{G_M^p}\right)^2 = \frac{F_2^n}{F_2^p} = \frac{1+4(d/u)}{4+(d/u)},
\quad Q^2\to\infty.
\]
Here $d$ and $u$ are the partonic density functions. The authors fit the data under the two
assumptions: $d/u=0$ and $d/u=0.2$. One more duality-motivated constraint is the equality
\[
\left(G_E^n/G_M^n\right)^2 = \left(G_E^p/G_M^p\right)^2
\]
applied for the highest $Q^2$ data points for the neutron electric form factor included into
the BBBA(07) fit. 

The GKex(05) model is in fact a modification of the QCD inspired vector dominance model (VDM)
by Gari and Kr\"uempelmann (GK) \cite{Gari:92} extended and fine-tuned by Lomon
\cite{Lomon:01,Lomon:02} in order to match the current and consistent earlier experimental data.
The data set used by Lomon includes the polarization transfer measurements,
which are directly related to the ratios of electric to magnetic form factors, and differential
cross section measurements of the magnetic form factors. The electric form factors derived from
the Rosenbluth separation of the differential cross section are only used for the lower range of
$Q^2$ where the magnetic contributions are less dominant. 
Among several versions of the parametrization considered by Lomon, we chose the latest one
``GKex(05)'' described in Ref.\ \cite{Lomon:06}. This version incorporates the data
that has become available since the publication \cite{Lomon:02} and has a bit better $\chi^2$.
The fitted parameters agree with the known constraints and the model is consistent with VDM at low $Q^2$,
while approaching perturbative QCD behavior at high $Q^2$. The quark-hadron duality constraint is not imposed.

\begin{figure*}[t]
\includegraphics[width=\linewidth]{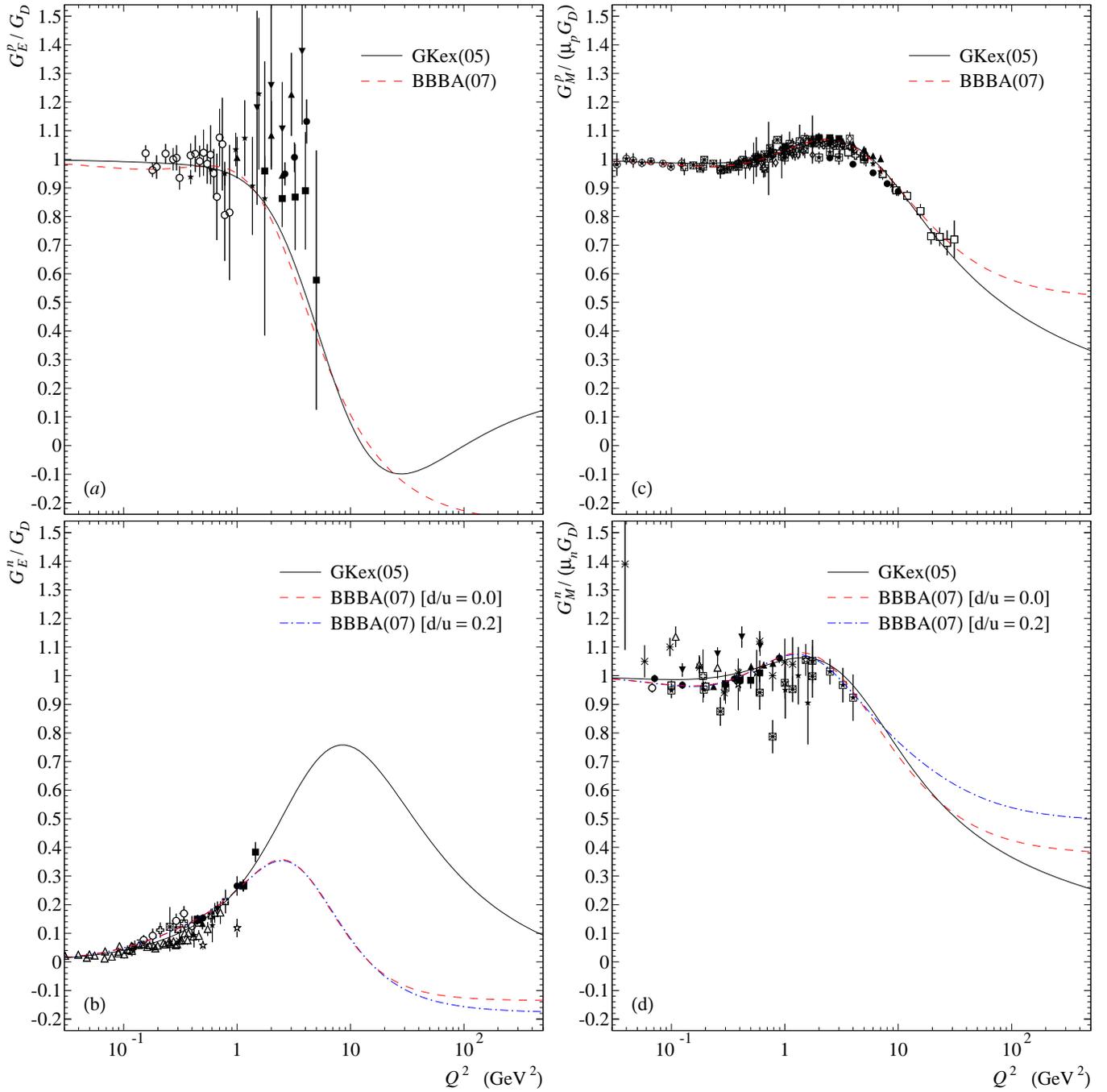}
\caption{Comparison of the GKex(05) and BBBA(07) models for the electric and magnetic form factors
         of proton and neutron (divided by the standard dipole $G_D$) with the data from electron
         scattering experiments. The data compilation is taken from 
         Refs.~\cite{Arrington:03a,Arrington:03b,Finn:04,Anderson:07,Day:07,Perdrisat:07}.
         The two versions of the BBBA(07) parametrization are shown for the neutron form factors.
       }
\label{Fig:GpE_GD+GpM_mGD+GnE_GD+GnM_mGD}
\end{figure*}

Figure \ref{Fig:GpE_GD+GpM_mGD+GnE_GD+GnM_mGD} shows a comparison of the GKex(05) and BBBA(07)
parametrizations for the form factors $G_E^{p,n}$ and $G_M^{p,n}$ divided by the standard
dipole $G_D$, against the experimental data extracted using either the Rosenbluth separation or
polarization transfer techniques (including a series of double-polarization measurements
of neutron knock-out from a polarized ${}^2\text{H}$ or ${}^3\text{He}$ targets).
The data assemblage is borrowed from Refs.\ \cite{Arrington:03a,Arrington:03b,Finn:04,Anderson:07}
and recent reviews \cite{Day:07,Perdrisat:07}.
It is seen from the figure that the models are numerically close to each other at low momentum
transfers covered by experiment, but diverge at high $Q^2$. The most serious disagreement between
the models is in the neutron electric form factor at $Q^2\gtrsim2~\text{GeV}^2$.
In section \ref{Statistical_analysis}, we examine how the model differences affect the extracted
value of the axial mass.

\subsection{Axial-vector and induced pseudoscalar form factors}

For the axial and pseudoscalar form factors we use the conventional
parametrizations \cite{LlewellynSmith:72}
\begin{align}
\label{F_A}
F_A(Q^2) & = F_A(0)\left(1+\frac{Q^2}{M^2_A}\right)^{-2}, \\
\label{F_P}
F_P(Q^2) & = \frac{2M^2}{m^2_\pi+Q^2}F_A(Q^2),
\end{align}
where $F_A(0)=g_A$ is the axial coupling,
$m_\pi$ is the charged pion mass,
and $M_A$ is the axial-vector mass treated as a free parameter.
In fact, Eq.\ \eqref{F_P} is a conjecture inspired by the hypothesis of partial conservation
of the axial current (PCAC), expectation that the form factor $F_P$ is dominated by the
pion pole near $Q^2=0$,
and the ``technical'' condition
\[
m_\pi^2\left|\frac{1}{F_A(0)}\frac{dF_A(Q^2)}{dQ^2}\right|_{Q^2=0}
=\frac{2m_\pi^2}{M_A^2} \ll 1,
\]
which is obviously fulfilled for the experimental lower limit of $M_A$.
Since the pseudoscalar contribution enters into the cross sections multiplied
by $(m_{\ell}/M)^2$, the uncertainty caused by this approximation may only be important
for $\nu_\tau/\overline\nu_\tau$ induced reactions (especially in the low-$Q^2$ range,
see, e.g., Refs.\ \cite{Hagiwara:04,Kuzmin:04}) and it is insignificant for reactions induced
by electron and muon (anti)neutrinos. 

\subsection{Constants}

The most precise determination of $V_{ud}$ comes from superallowed nuclear
beta decays ($0^+\to0^+$ transitions). We adopt the weighted average of the nine
best measured superallowed decays  
$V_{ud}^{\text{(SA)}}=0.97377\pm0.00027$ recommended by the Particle Data Group
(PDG)~\cite{Yao:06}.
Note that this value is consistent with that of the PIBETA experiment at
PSI \cite{Pocanic:04}, $V_{ud}^{\text{(PIBETA)}}=0.9728\pm0.0030$, obtained
from the measured branching ratio for pion beta decay $\pi^+\to\pi^0e^+\nu$.

For the axial-vector and Fermi coupling constants, we use the standard PDG
averaged values: $g_A=-1.2695\pm0.0029$ and $G_F=1.16637\times10^{-5}~\text{GeV}^2$
\cite{Yao:06}. In several papers (see, e.g., Ref.\ \cite{Nakamura:02} and references therein)
it is suggested to use the value $G_F'=1.1803\times10^{-5}~\text{GeV}^2$ obtained from
$0^+\to0^+$ nuclear $\beta$ decays, rather than the standard $G_F$ obtained from
muon $\beta$ decay. The coupling constant $G_F'$ subsumes the bulk of the inner radiative
corrections. However, some neutrino experiments already take the radiative corrections into
account (sometimes  in quite different ways) in the measured cross sections.
That is why, in this study, we simply add the corresponding difference (of about 2\%)
to the overall uncertainty of the fit. Note that using the $G_F'$ instead of $G_F$
would lead to a few percent \emph{decrease} of the output value of $M_A$.

\section{Relativistic Fermi gas model}

Since the main part of the experimental data on the QES cross sections for nuclear targets
was not corrected for nuclear effects, we must take these into account in our calculations.
In the present work, we use the RFG model by Smith and Moniz \cite{Smith:72} incorporated
as a standard tool into essentially all neutrino event generators employed in accelerator
and astroparticle neutrino experiments.

According to RFG, the hadronic tensor $W_{\alpha\beta}$ given by Eq.~\eqref{W}
must be replaced with the tensor $T_{\alpha\beta}$, which describes the bound nucleon.
This tensor is of the same Lorentz structure as $W_{\alpha\beta}$ and is defined by
the six invariant nuclear structure functions $T_i(Q^2)$.
Thus, in the in the lab.\ frame 
\begin{align}\label{W_nucl}
T_{\alpha\beta}\left(p_{\text{lab}},q\right) 
= & -g_{\alpha\beta} T_1+g_{0\alpha} g_{0\beta} T_2           \nonumber \\
  &  -\frac{i\varepsilon_{\alpha\beta 0\delta} q^\delta}{2M_t} T_3 
     +\frac{q_\alpha q_\beta}{M_t^2} T_4                      \nonumber \\
  &  +\frac{g_{0\alpha}q_\beta+q_\alpha g_{0\beta}}{2M_t}T_5  \nonumber \\
  &  +i\frac{g_{0\alpha}q_\beta-q_\alpha g_{0\beta}}{2M_t}T_6 \nonumber \\
= & \int d\vec{p}f(\vec{p},\vec{q})W_{\alpha\beta}(p,q),
\end{align}
where $p_{\text{lab}}=(M_t,\vec{0})$, $M_t$ is the mass of the target nucleus, and
\[
f(\vec{p},\vec{q})=v_{\text{rel}}^{-1}\;\overline{n}_i(\vec{p})\left[1-n_f(\vec{p}+\vec{q})\right].
\]
The function $\overline{n}_i(\vec{p})$ is the Fermi momentum distribution of the target nucleons,
satisfying the normalization condition
\begin{equation*}
\int \overline{n}_i(\vec{p})d\vec{p}=1.
\end{equation*}
The factor $1-n_f(\vec{p}+\vec{q})$ (the unoccupation probability) takes into account the Pauli
blocking for the outgoing nucleon. 
The relative velocity $v_{\text{rel}}$ which represents the flux of incident particles,
is given by
\[
v_{\text{rel}} = |(kp)|/(E_{\nu}M_t).
\]
Explicitly defining the three-momenta $\vec{q}$, $\vec{p}$, and $\vec{p}$,
\begin{align*}
\vec{q} & = \left(0,0,|\vec{q}|\right),           \\
\vec{p} & = \left(\sin\theta_{\vec{k}},0,
            \cos\theta_{\vec{k}}\right)|\vec{q}|, \\
\vec{p} & = \left(\sin\theta_{\vec{p}}\cos\phi_{\vec{p}},
                  \sin\theta_{\vec{p}}\sin\phi_{\vec{p}},
                  \cos\theta_{\vec{p}}\right)|\vec{p}|,
\end{align*}
one obtains 
\[
v_{\text{rel}} = \left[E_{\vec{p}}-|\vec{p}|\left(\cos\theta_{\vec{k}}\cos\theta_{\vec{p}}
      +\sin\theta_{\vec{k}}\sin\theta_{\vec{p}}\sin\varphi_{\vec{p}}\right)\right]/M_t,
\]
where
\[
E_{\vec{p}}=\sqrt{\vec{p}^2+M^2}-\epsilon_b
\]
is the total energy of the bound nucleon and $\epsilon_b$ is the effective binding energy.
The angle $\theta_{\vec{k}}$ is defined by
\[
\cos\theta_{\vec{k}}=\frac{E_\nu^2+\vec{q}^2+m_\ell^2}{2E_\nu|\vec{q}|}.
\]
For determining the angle $\theta_{\vec{p}}$, one can use the energy conservation law
defined by delta-function
\begin{gather*}
\delta(E_{\vec{p}}-E_{\vec{p}+\vec{q}}+\nu) = 
\frac{1}{2|\vec{p}||\vec{q}|}\delta\left(\cos\theta_{\vec{p}}-\cos\theta_{\vec{p}}^{\,0}\right),
\end{gather*}
where $\nu=E_\nu-E_\ell$ and
\[
E_{\vec{p}+\vec{q}}=\sqrt{\vec{p}^2+\vec{q}^2+2|\vec{p}||\vec{q}|\cos\theta_{\vec{p}}+M^2}.
\]
is the total energy of the outgoing nucleon. Then the condition
\[
\cos\theta_{\vec{p}}=\cos\theta_{\vec{p}}^{\,0}
=\frac{(\nu+E_{\vec{p}})^2-(E_{\vec{p}}+\epsilon_b)^2-\vec{q}^2}{2|\vec{p}||\vec{q}|}
\]
must be obeyed.

The nuclear structure functions are the linear combination of the $W_i$
and can be straightforwardly calculated from Eqs.\ \eqref{W} and \eqref{W_nucl}:
\begin{align*}
T_1 = & a_1W_1
        +\frac{1}{2}\left(a_2-a_3\right) W_2,                                 \\
T_2 = &  \left[\frac{\vec{q}^2-\nu^2}{2\vec{q}^2}\left(a_2-a_3\right) 
        +\frac{\nu^2}{\vec{q}^2}a_3+a_4-\frac{2\nu}{|\vec{q}|}a_5\right]W_2,  \\
T_3 = &  \frac{M_t}{M}\left(a_7-\frac{\nu}{|\vec{q}|}a_6 \right)W_3,          \\
T_4 = &  \frac{M_t^2}{M^2}\left[\frac{M^2}{2\vec{q}^2}
         \left(3a_3-a_2\right)W_2+a_1W_4+\frac{M}{|\vec{q}|}a_6W_5\right],    \\
T_5 = &  \frac{M_t}{|\vec{q}|}\left[\frac{\nu}{|\vec{q}|}
         \left(a_2-3a_3\right)+2a_5\right]W_2                                 \\
      & +\frac{M_t}{M}\left(a_7-\frac{\nu}{|\vec{q}|}a_6\right)W_5,           \\
T_6 = &  \frac{M_t}{M}\left(a_7-\frac{\nu}{|\vec{q}|}a_6\right)W_6.
\end{align*}
The coefficients $a_i$ are given by
\begin{align*}
a_1 = &               \int f(\vec{p},\vec{q})d\vec{p},                                \\
a_2 = & \frac{1}{M^2} \int f(\vec{p},\vec{q})\vec{p}^2d\vec{p},                       \\
a_3 = & \frac{1}{M^2} \int f(\vec{p},\vec{q})\vec{p}^2\cos^2\theta_{\vec{p}}d\vec{p}, \\
a_4 = & \frac{1}{M^2} \int f(\vec{p},\vec{q})E_{\vec{p}}^2 d\vec{p},                  \\
a_5 = & \frac{1}{M^2} \int f(\vec{p},\vec{q})
                                    E_{\vec{p}}|\vec{p}|\cos\theta_{\vec{p}}d\vec{p}, \\
a_6 = & \frac{1}{M}   \int f(\vec{p},\vec{q})|\vec{p}|\cos\theta_{\vec{p}}d\vec{p},   \\
a_7 = & \frac{1}{M}   \int f(\vec{p},\vec{q})E_{\vec{p}}d\vec{p}.
\end{align*}
\begin{table}[hbt]
\centering
\caption{Proton and neutron Fermi momenta and binding energies (in MeV)
         for selected nuclei.
\label{table:RFGparameters}
\hfill}
\begin{tabular}{c@{\qquad\quad}c@{\qquad\qquad}c@{\qquad\qquad}c@{\qquad\qquad}c}
  \hline\noalign{\smallskip}
     Nucleus         &$p_F^p$&$\epsilon_b^p$&$p_F^n$&$\epsilon_b^n$ \\
  \hline\noalign{\smallskip}
  ${}_{~~6}^{~12}$C~ &  221  &     25.6     &  221  &     25.6      \\  
  \noalign{\smallskip}
  ${}_{~~7}^{~14}$N~ &  223  &     26.2     &  223  &     26.1      \\  
  \noalign{\smallskip}
  ${}_{~~8}^{~16}$O~ &  225  &     26.6     &  225  &     26.6      \\  
  \noalign{\smallskip}
  ${}_{~~9}^{~19}$F~ &  233  &     28.4     &  233  &     28.3      \\  
  \noalign{\smallskip}
  ${}_{~10}^{~20}$Ne &  230  &     27.8     &  230  &     27.8      \\  
  \noalign{\smallskip}
  ${}_{~13}^{~27}$Al &  239  &     29.5     &  239  &     29.4      \\  
  \noalign{\smallskip}
  ${}_{~18}^{~40}$Ar &  242  &     30.7     &  259  &     35.0      \\  
  \noalign{\smallskip}
  ${}_{~26}^{~56}$Fe &  251  &     33.0     &  263  &     36.1      \\  
  \noalign{\smallskip}
  ${}_{~35}^{~80}$Br &  245  &     31.5     &  270  &     38.1      \\  
  \noalign{\smallskip}
  \hline\noalign{\smallskip}
\end{tabular}
\end{table}
Finally, in order to describe the neutrino scattering off a bound nucleon, one
should substitute $M \longmapsto M_t$ and $W_i \longmapsto T_i$ in Eq.~\eqref{R};
then the differential cross-section can be calculated according to Eq.~\eqref{cs}
(see Ref.\ \cite{Lyubushkin:08a} for more details).
Table \ref{table:RFGparameters} collects the values of proton and neutron Fermi momenta
$p_F^{p,n}$ and binding energies $\epsilon_b^{p,n}$ for several nuclei, used in our
numerical calculations.

\section{Statistical analysis of the data}
\label{Statistical_analysis}

\subsection{Description of experimental data}

We have examined and classified all available experimental data on quasielastic scattering
with $\Delta Y=0$. Published results from the relevant experiments with $\nu_\mu$ and
$\overline\nu_\mu$ beams from accelerators at
ANL  \cite{Novey:67,Kustom:69a,Mann:72,Mann:73,S.Barish:75,Singer:77,%
           S.Barish:77a,Miller:82},
BNL  \cite{Cazzoli:76,Cnops:78a,Fanourakis:80,Baker:81b,Kitagaki:86a,%
           Kitagaki:86b,Abe:86,Ahrens:87,Ahrens:88,Kitagaki:90,Sakuda:03,Furuno:03},
FNAL \cite{Asratyan:82,Kitagaki:83,Asratyan:84a,Asratyan:84b,Ammosov:86,Ammosov:87b,Suwonjandee:04},
CERN \cite{Burmeister:65,Franzinetti:66,Young:67,Orkin-Lecourtois:67,%
           Holder:68,Budagov:69c,Eichten:72,Eichten:73a,Sciulli:74a,%
           Haguenauer:74,Rollier:75,Deden:75,Bonetti:77,Erriquez:77,%
           Rollier:78,Dewit:78,Erriquez:78a,Pohl:79b,Armenise:79a,%
           Allasia:90,Petti:04,Lyubushkin:06,Martinez:07,Lyubushkin:08a}, and
IHEP \cite{Makeev:81,Belikov:81,Belikov:82a,Belikov:82b,Belikov:83,%
           Belikov:85a,Belikov:85b,Grabosch:86b,Grabosch:88,Brunner:90}
are included dating from the end of sixties to the present day, covering
a variety of nuclear targets, with energies ranging from about $150$~MeV (ANL experiments)
to about $350$~GeV (NuTeV).
Pertinent additional information was borrowed from the review articles and data compilations
\cite{Perkins:72,Derrick:74,Perkins:75,Cline:77,Wachsmuth:77,Ermolov:78,Musset:78,Alekhin:87,%
      Sakuda:02,Ammosov:92,Baltay:94,Zeller:03,Fleming:06,Sorel:07,Gran:07}.

All the fits are done with the CERN function minimization and error analysis package ``MINUIT''
(version 94.1)~\cite{MINUIT}, taking care of getting an accurate error matrix.
The errors of the output parameters quoted below correspond to the usual one-standard-deviation
($1\sigma$) errors (MINUIT default).

For the analysis, we have selected the most statistically reliable measurements of the total
and differential cross sections for each nuclear target, which were not superseded or reconsidered
(due to increased statistics, revised normalization, etc.) in the posterior reports of the same experimental
groups.
Finally, we include into the global fit the data on the total cross sections from
Refs.~\cite{S.Barish:77a,Fanourakis:80,Baker:81b,Kitagaki:83,Ammosov:87b,Suwonjandee:04,Young:67,%
            Bonetti:77,Pohl:79b,Martinez:07,Lyubushkin:08a,Belikov:85b,Brunner:90}
and the data for the differential cross sections from
Refs.~\cite{Bonetti:77,Allasia:90,Belikov:82a,Belikov:82b,Belikov:85b,Brunner:90,Musset:78}.
The remaining data are either obsolete, or exhibit uncontrollable systematic errors and/or fall
well outside the most probable range determined through the fit of the \emph{full} dataset;
the value of $\chi^2$ evaluated for each subset of the rejected data usually exceeds $(3-4)~\text{NDF}$.

Since the differential cross sections $d\sigma/dQ^2$
were measured, as a rule, within rather wide ranges of the energy spectra of $\nu_\mu$ and
$\overline\nu_\mu$ beams, we use only the data from such experiments, in which the spectra
were known (measured or calculated and then calibrated) with reasonably good accuracy.
All the energy spectra
(borrowed from Refs.\ \cite{Bonetti:77,Armenise:79a,Belikov:85b,Musset:78,Ammosov:92,Budagov:69a,Allasia:84})
necessary for numerical averaging of the calculated differential cross sections and distributions
were parametrized. To avoid the loss of accuracy, the precision of these parametrizations was chosen
to be at least an order of magnitude better than the experimental accuracy of the spectra themselves.
For a verification, we have estimated the mean energies of the beams for different energy intervals,
and have compared these against the published values.

\begin{figure*}[t]
\centering
\includegraphics[width=\linewidth]{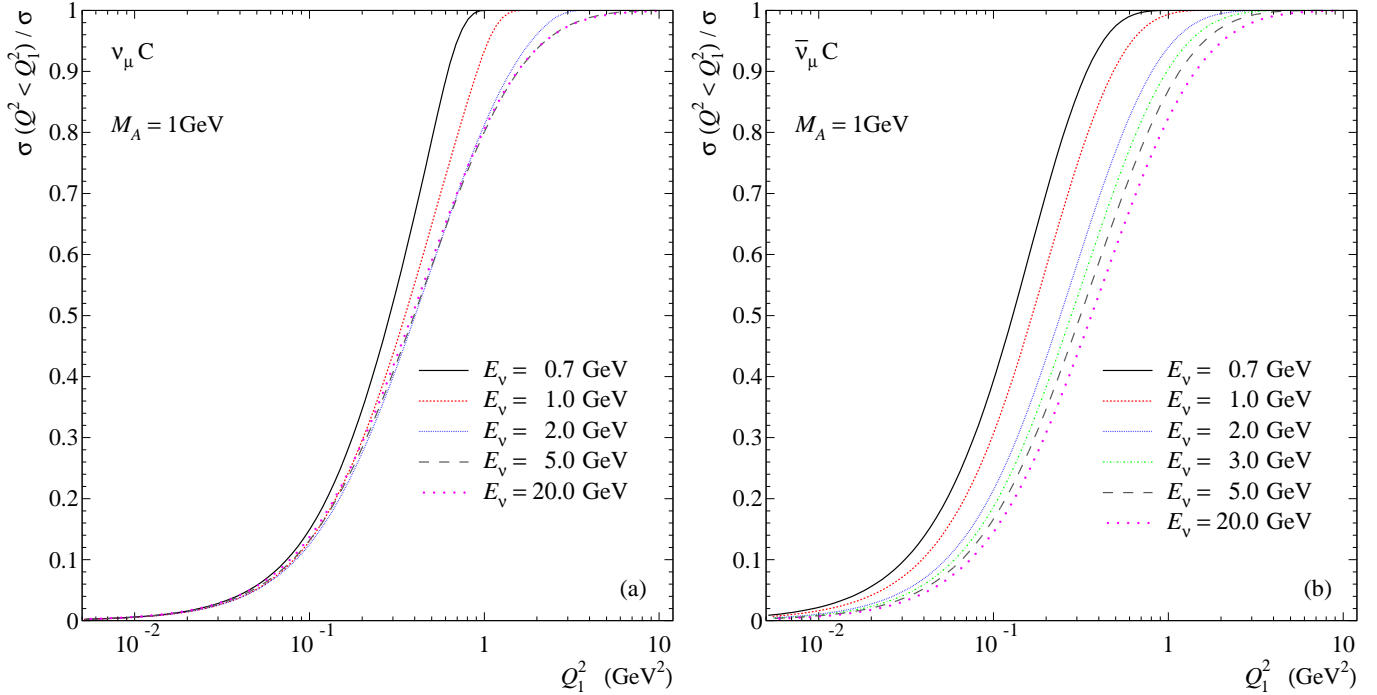}
\caption{The ratio $R=\sigma\left(Q^2<Q_1^2\right)/\sigma$ vs.\ $Q_1^2$,
         evaluated for $\nu_\mu$ and $\overline\nu_\nu$ quasielastic
        interactions with carbon target at several (anti)neutrino energies.
         The $M_A$ value is taken to be 1~GeV.
        }
\label{Fig:RQES(Q2)_BBBA25_C}
\end{figure*}

\begin{figure*}[t]
\centering
\includegraphics[width=0.91\linewidth]{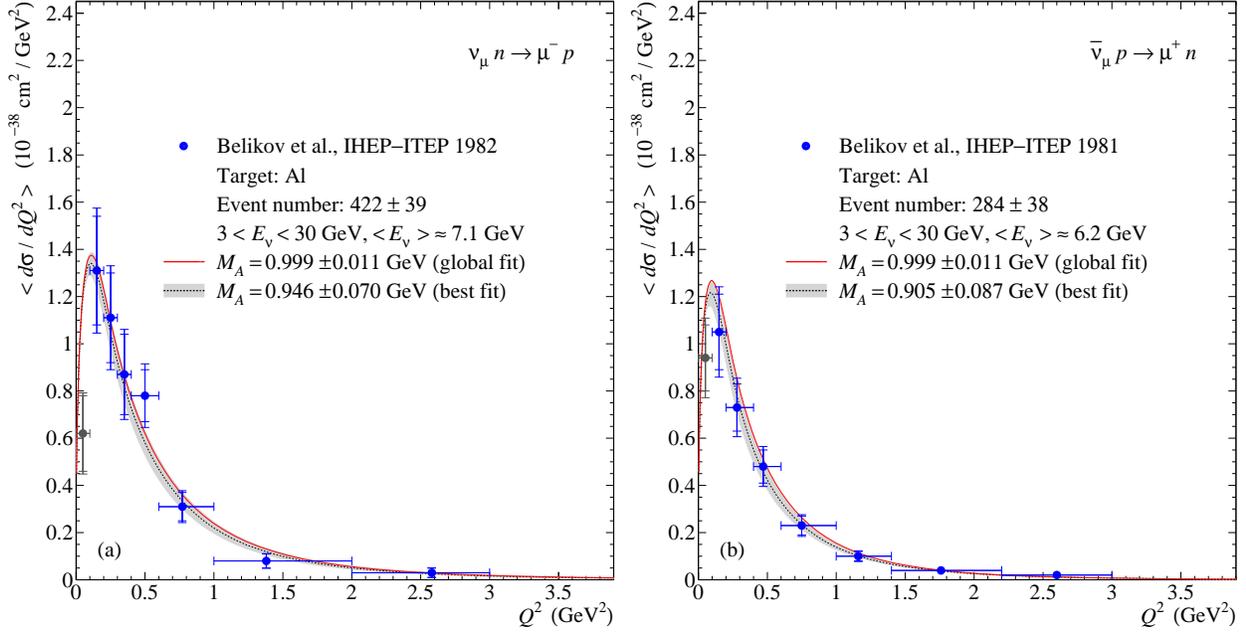}
\caption{Flux-weighted differential cross sections for
         $\nu_{\mu}n\to\mu^-p$ (a) and $\overline\nu_{\mu}p\to\mu^+n$ (b)
         measured in the IHEP-ITEP experiment with a spark chamber detector
         with aluminium filters and exposed to the U70 broad-band $\nu_{\mu}$
         and $\overline\nu_{\mu}$ beams of the Serpukhov PS \cite{Belikov:81,Belikov:82b}.
         The inner and outer bars indicate statistical and total errors,
         respectively; the overall systematic error of about 10\% is due mainly
         to uncertainties of the flux normalization and scanning/triggering
         efficiencies.
         The curves are the calculated cross sections averaged over the
         experimental $\nu_{\mu}$ and $\overline\nu_{\mu}$ energy spectra
         borrowed from Refs.\ \cite{Belikov:85b,Ammosov:92}.
         The energy range and estimated mean energies are given in the legends.
         The dashed curves are for the best fit to the IHEP-ITEP data, while
         the solid curves correspond to the global fit to all QES data.
         The points shown by grey symbols are excluded from the fits (see text).
         Shaded bands represent $1\sigma$ deviations from the best-fitted values
         of $M_A$ given in the legends.
        }
\label{Fig:dsQES_dQ2_Belikov_IHEP-ITEP81,82}
\end{figure*}

\begin{figure*}[t]
\centering
\includegraphics[width=\linewidth]{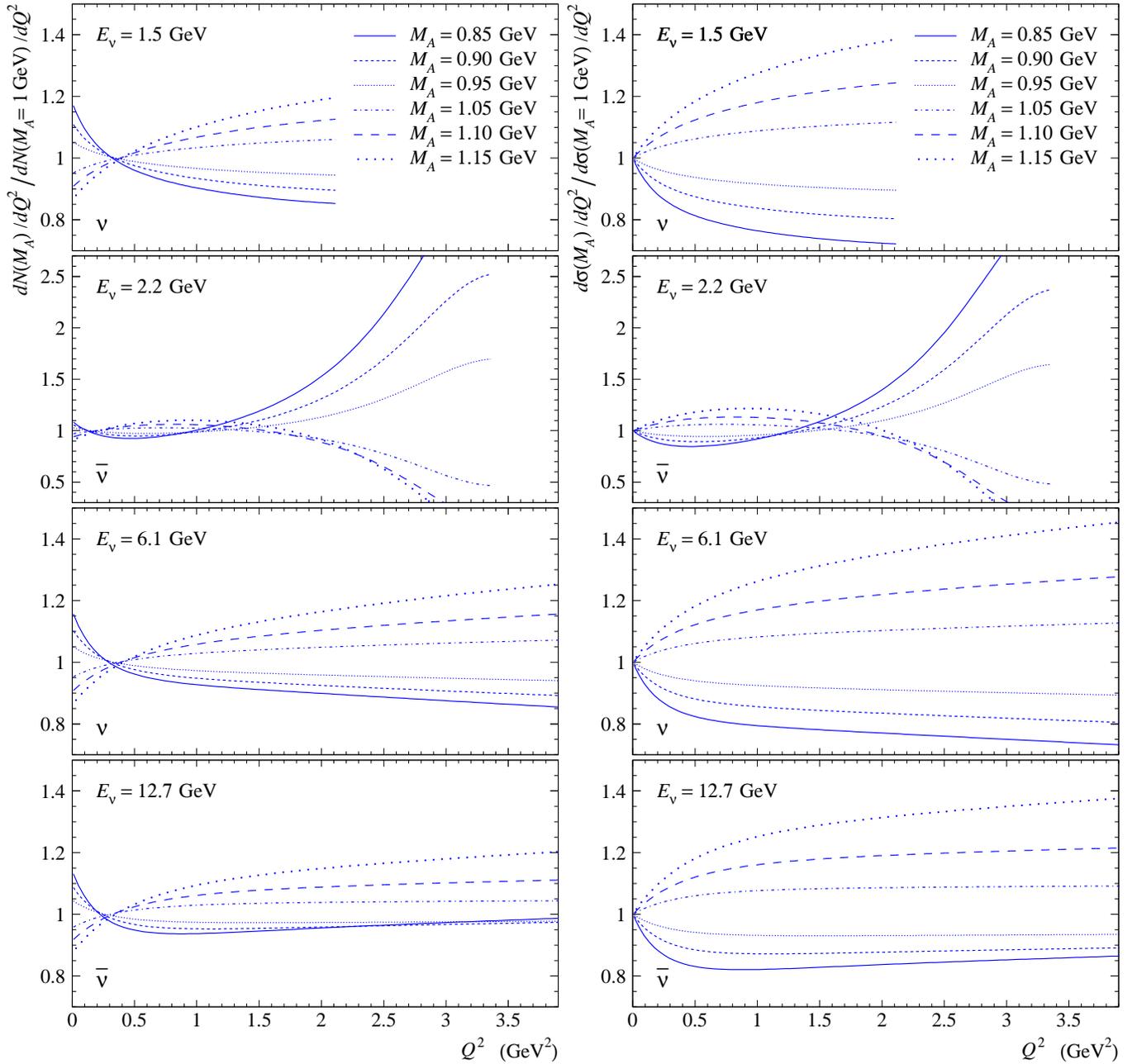}
\caption{The distributions $dN/dQ^2$ and differential cross sections $d\sigma/dQ^2$
         vs.\ $Q^2$ for $\nu_{\mu}n$ and $\overline\nu_{\mu}p$ quasielastic scattering,
         calculated with different $M_A =0.85$, $0.90$, $0.95$, $1.05$, $1.10$, and $1.15$~GeV
         and normalized to the corresponding quantities
         calculated with $M_A=1~\text{GeV}$ at four fixed values of energy corresponding
         to the mean (anti)neutrino beam energies in experiments 
         HLBC~1969 \cite{Budagov:69c},
         Gargamelle~1979 \cite{Armenise:79a},
         SKAT~1981 \cite{Makeev:81}, and
         FNAL~1984 \cite{Asratyan:84a,Asratyan:84b}
         (see Figs.~\protect\ref{Fig:dNQES_dQ2_Budagov_HLBC69_1}--\protect\ref{Fig:dNQES_dQ2_Asratyan_FNAL84_1}
         below).
         The curves in the four upper panels end up at the kinematical boundaries.
        }
\label{Fig:Ratios}
\end{figure*}

\begin{figure*}[t]
\centering
\includegraphics[width=\linewidth]{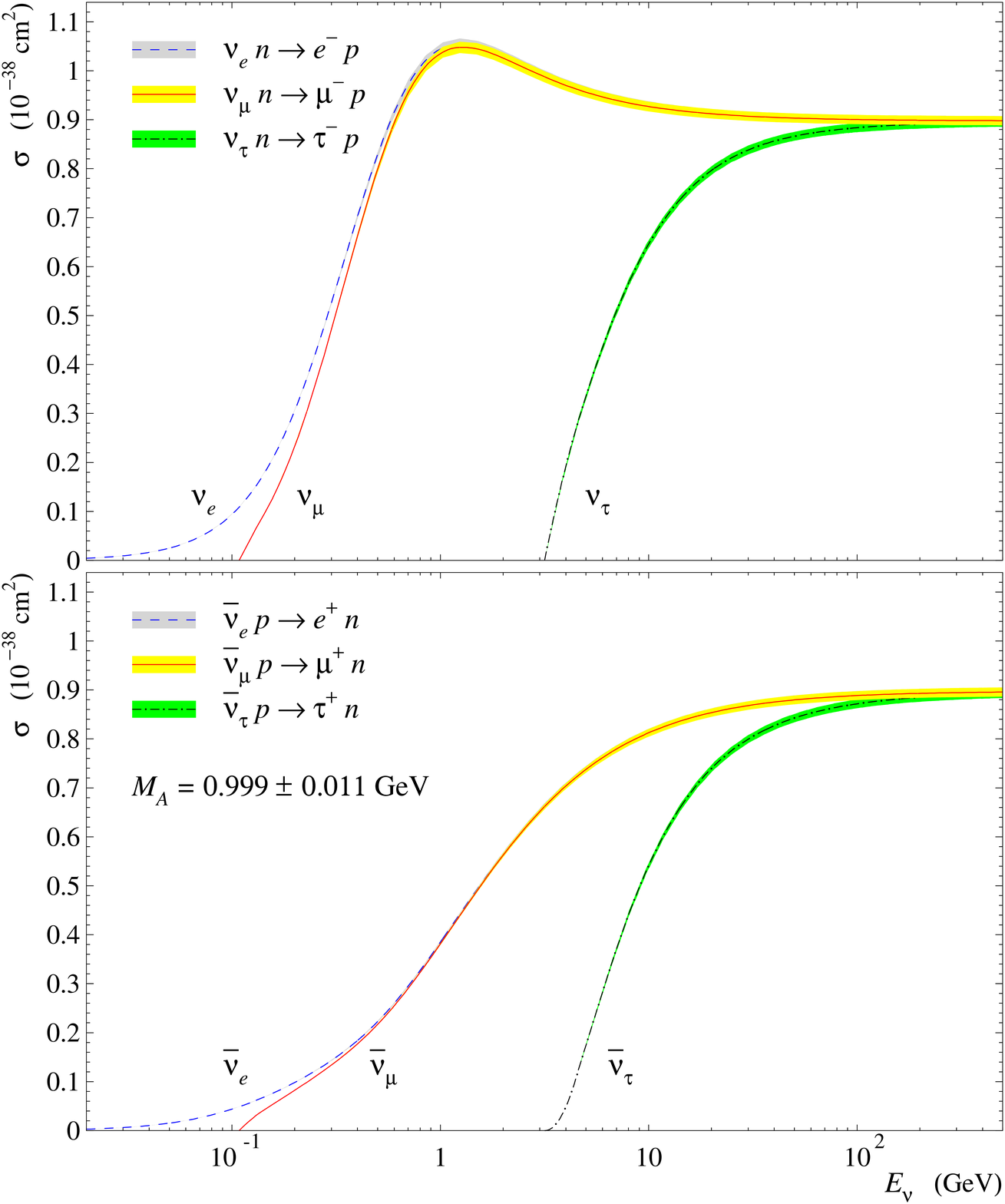}
\caption{Total quasielastic cross sections for electron, muon and $\tau$ neutrino and
        antineutrino interactions with free nucleons calculated with the best-fit value
        of $M_A=0.999\pm0.011~\text{GeV}$ using the BBBA(07) vector form factors.
        Shaded bands represent the uncertainty due to the $1\sigma$ error in $M_A$.
        }
\label{Fig:sQES_emt_101.0.00.301.01_1_BBBA25}
\end{figure*}

The analyses were performed for neutrino and antineutrino data separately, and for the full set of
the $\nu$ and $\overline\nu$ data together. For each fit, we have included the data for either total
or differential cross sections, as well as for the cross sections of both types together.
The main results of the analysis are summarised in Tables \ref{table:M_A} and \ref{table:M_A_Bodek}
and illustrated in
Figs.\ \ref{Fig:sQES_101.0.00.301.01_1_BBBA25_NT_PART1_PRD}--\ref{Fig:sQES_K2K06_101.0.00.301.01_1_BBBA25_SM_PRD}.
Let us discuss these results in details.

\subsection{Main results of the global fit}

As is seen from Table \ref{table:M_A}, the differences between the values of $M_A$ extracted from the fits
of each type, performed with the BBBA(07) and GKex(05) models for the vector form factors vary between
0.3\% and 1.3\% that is less than or of the order of one standard deviation in the $M_A$ extractions
and is comparable with the accuracy of the most precise measurements of the electric and magnetic
form factors. The values of $\chi^2/\text{NDF}$ are essentially the same for BBBA(07) and GKex(05).
The differences in the $M_A$ values obtained with the two versions of the BBBA(07) model corresponding to
$d/u=0$ and $0.2$ (the latter is not shown in the table) are less than $0.2\%$ that is practically negligible.
Therefore, in the following we will solely discuss the $d/u=0$ case. 
\begin{table*}[htb]
\centering
\caption{\label{table:M_A}
         Values of $M_A$ (given in GeV), extracted by fitting the $\nu_\mu$, $\overline\nu_\mu$, and
         $\nu_\mu+\overline\nu_\mu$ data on total and differential QES cross sections, using the BBBA(07) and GKex(05)
         models for the vector form factors of the nucleon.
         The $\chi^2/\text{NDF}$ values for each fit are shown in parentheses.
\hfill}
\begin{tabular}{ccc@{\qquad\qquad}ccc}
\hline\noalign{\smallskip}
\multicolumn{3}{c}{\bf BBBA(07)}                                  &  \multicolumn{3}{c}{\bf GKex(05)}                                 \\
\noalign{\smallskip}
\hline\noalign{\smallskip}
$M_A^\nu$       & $M_A^{\overline\nu}$ & $M_A^{\nu+\overline\nu}$ & $M_A^\nu$       & $M_A^{\overline\nu}$ & $M_A^{\nu,\overline\nu}$ \\
\noalign{\smallskip}
\hline\noalign{\smallskip}
\multicolumn{6}{l}{\bf Fit to the total cross sections:}                                                                              \\
\noalign{\smallskip}
$0.994\pm0.017$ & $1.047\pm0.025$      & $1.011\pm0.014$          & $0.986\pm0.017$ & $1.035\pm0.025$      & $1.001\pm0.014$          \\
$(83/82)$       & $(134/62)$           & $(220/145)$              & $(83/82)$       & $(137/62)$           & $(222/145)$              \\
\noalign{\smallskip}
\hline\noalign{\smallskip}
\multicolumn{6}{l}{\bf Fit to the differential cross sections:}                                                                       \\
\noalign{\smallskip}
\noalign{\smallskip}
$0.979\pm0.020$ & $0.991\pm0.029$      & $0.983\pm0.017$          & $0.976\pm0.020$ & $0.982\pm0.030$      & $0.978\pm0.017$          \\
$(45/48)$       & $(26/37)$            & $(71/86)$                & $(45/48)$       & $(25/37)$            & $(70/86)$                \\
\noalign{\smallskip}
\hline\noalign{\smallskip}
\multicolumn{6}{l}{\bf Fit to the total and differential cross sections:}                                                             \\
\noalign{\smallskip}
$0.988\pm0.013$ & $1.023\pm0.018$      & $0.999\pm0.011$          & $0.981\pm0.013$ & $1.012\pm0.019$      & $0.991\pm0.011$          \\
$(128/131)$     & $(163/100)$          & $(293/232)$              & $(128/131)$     & $(163/100)$          & $(293/232)$              \\
\noalign{\smallskip}
\hline\noalign{\smallskip}
\end{tabular}
\end{table*}

The $M_A$ values obtained from the fits to the differential cross sections are systematically lower those obtained
from the total cross sections. The differences amount $\sim1.5\%$ ($\sim5.7\%$) for $\nu_{\mu}$ ($\overline{\nu}_{\mu}$)
that is (especially in antineutrino case) above the statistical error of the fit and is caused mainly by uncertainties
in the energy spectra of $\nu_{\mu}$ and $\overline{\nu}_{\mu}$ and, in lesser extent, in the nuclear effects.

\begin{figure*}[t]
\includegraphics[width=\linewidth]{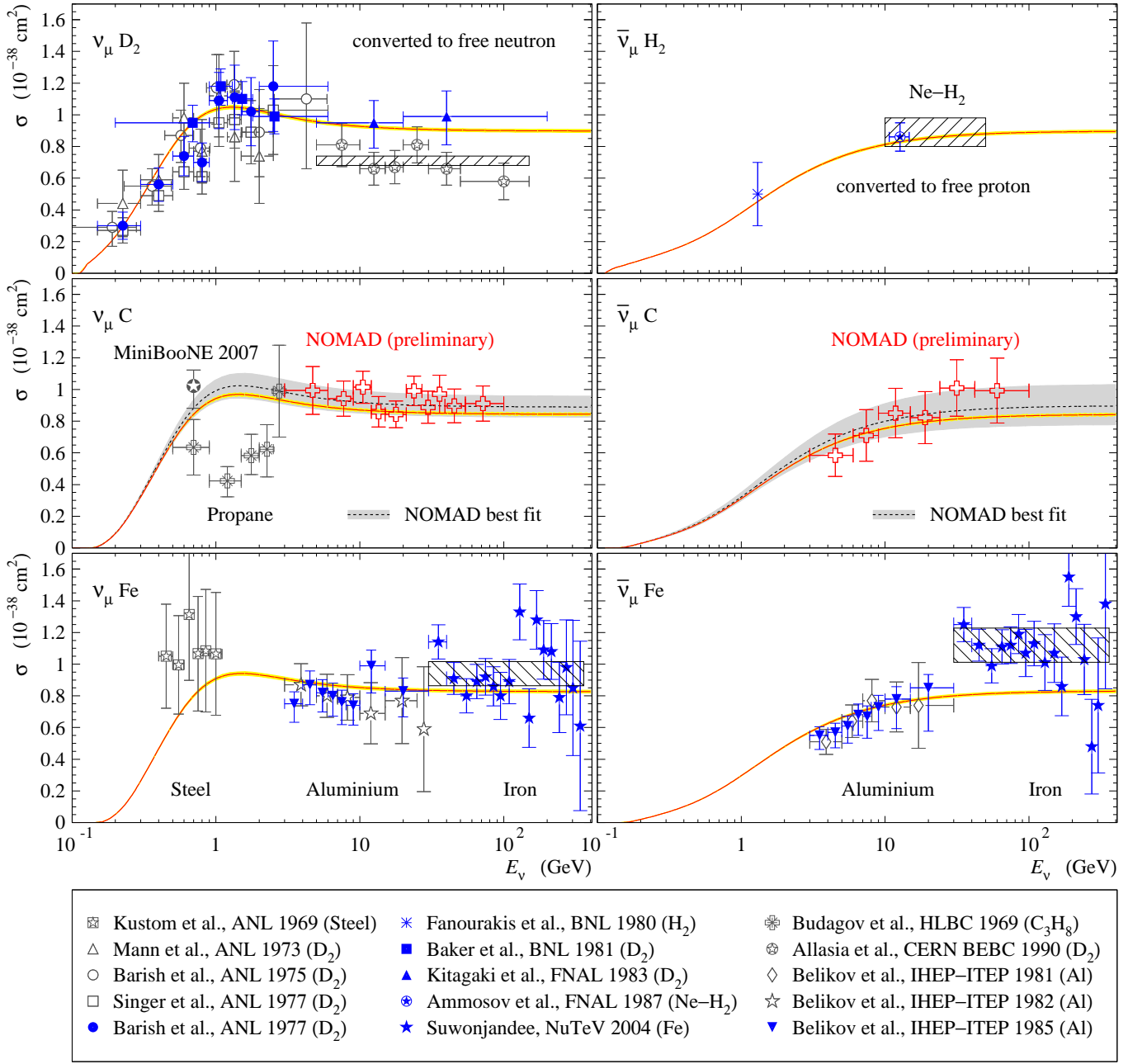}
\caption{Total quasielastic $\nu_{\mu}n$ and $\overline{\nu}_{\mu}p$ cross sections
         measured in experiments with deuterium, hydrogen, carbon/propane, aluminium, and iron/steel
         targets at
         ANL 1969 \cite{Kustom:69a},
         ANL 1973 \cite{Mann:73},
         ANL 1975 \cite{S.Barish:75}, 
         ANL 1977 \cite{Singer:77,S.Barish:77a},
         BNL 1980 \cite{Fanourakis:80},
         BNL 1981 \cite{Baker:81b},
         FNAL 1983 \cite{Kitagaki:83},
         FNAL E180 1984 \cite{Asratyan:84a,Asratyan:84b} (rectangle in top right panel),
         FNAL E180 1987 \cite{Ammosov:87b},
         NuTeV 2004 \cite{Suwonjandee:04} (points and rectangles in bottom panels),
         CERN HLBC 1969 \cite{Budagov:69c},
         CERN BEBC 1990 \cite{Allasia:90} (points and rectangle in top left panel),
         CERN NOMAD 2008 \cite{Lyubushkin:08a} (preliminary),
         IHEP-ITEP 1981 \cite{Belikov:81},
         IHEP-ITEP 1982 \cite{Belikov:82b}, and
         IHEP-ITEP 1985 \cite{Belikov:85a,Belikov:85b}.
         The deuterium and neon-hydrogen data were converted to a free neutron/proton target
         by the authors of the experiments.
         The MiniBooNE 2007 point \cite{Aguilar-Arevalo:07} recalculated from the reported
         value of $M_A=1.23\pm0.20~\text{GeV}$ is also shown for comparison.
         The error bars represent the total errors which include the flux normalization uncertainties.
         The solid curves and narrow shaded bands are calculated with the BBBA(07) model
         for the vector form factors, with $M_A=0.999\pm0.011~\text{GeV}$, the value
         obtained from the global fit to a subset of the full data set of total and
         differential cross sections (233 data points).
         The points shown by grey symbols
         are excluded from the fit, being either superseded by newer experiments,
         or not satisfying our selection criteria.
         The dashed curves and corresponding bands are the cross sections obtained by
         fitting the NOMAD 2008 alone with the GKex(2005) vector form factors
        (separately for  $\nu_{\mu}$ and $\overline{\nu}_{\mu}$ data).
       }
\label{Fig:sQES_101.0.00.301.01_1_BBBA25_NT_PART1_PRD}
\end{figure*}

\begin{figure*}[t]
\includegraphics[width=\linewidth]{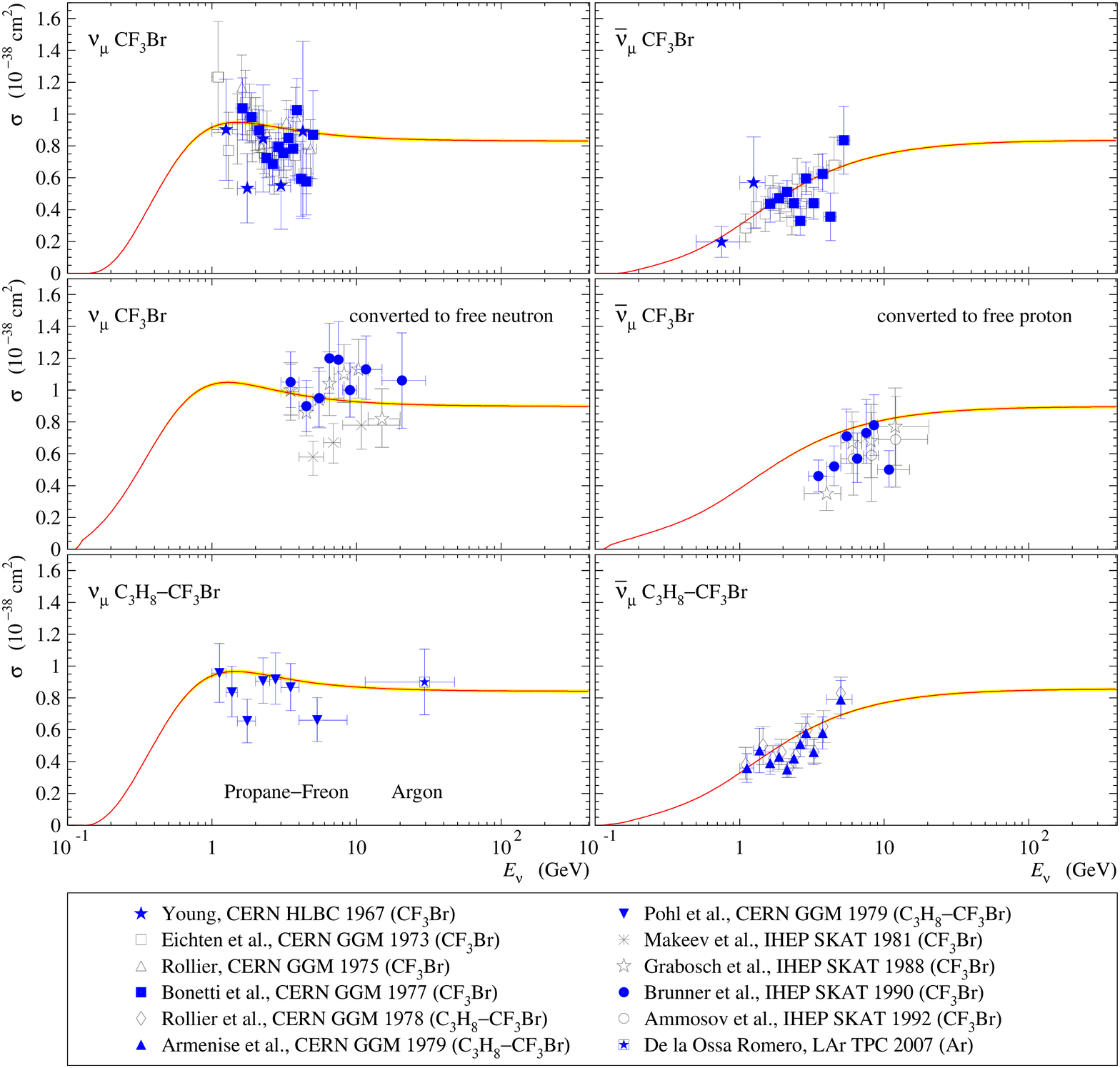}
\caption{Total quasielastic $\nu_{\mu}n$ and $\overline{\nu}_{\mu}p$ cross sections
         measured with the freon and propane-freon filled bubble chamber experiments
         CERN HLBC 1966 \cite{Franzinetti:66},
         CERN HLBC 1967 \cite{Young:67},
         CERN GGM  1973 \cite{Eichten:73a},
         CERN GGM  1975 \cite{Rollier:75,Perkins:75},
         CERN GGM  1977 \cite{Bonetti:77},
         CERN GGM  1978 \cite{Rollier:78},
         CERN GGM  1979 \cite{Armenise:79a,Pohl:79b},
         IHEP SKAT 1981 \cite{Makeev:81},
         IHEP SKAT 1988 \cite{Grabosch:88},
         IHEP SKAT 1990 \cite{Brunner:90}, and
         IHEP SKAT 1992 \cite{Ammosov:92}.
         The point recently obtained in experiment with the Liquid Argon Time Projection Chamber
         (LAr TPC 2007) \cite{Martinez:07}
         is also shown.
         The SKAT datapoints were converted from freon to a free neutron/proton target by the authors
         of the experiments.
         The error bars represent the total errors which include the uncertainties due to flux
         normalization and nuclear Monte Carlo.
         The solid curves and narrow shaded bands are calculated with the BBBA(07) model
         for the vector form factors, with the global fit value of $M_A=0.999\pm0.011~\text{GeV}$.
         See caption of Fig.\ \protect\ref{Fig:sQES_101.0.00.301.01_1_BBBA25_NT_PART1_PRD} for
         more details.
       }
\label{Fig:sQES_101.0.00.301.01_1_BBBA25_NT_PART2_PRD}
\end{figure*}

Figures \ref{Fig:sQES_101.0.00.301.01_1_BBBA25_NT_PART1_PRD} and
        \ref{Fig:sQES_101.0.00.301.01_1_BBBA25_NT_PART2_PRD}
show a compilation of the available data on the total QES cross sections for the following
nuclear targets:
hydrogen 
\cite{Fanourakis:80},
deuterium
\cite{Mann:73,S.Barish:75,Singer:77,S.Barish:77a,Baker:81b,Kitagaki:83,Allasia:90},
carbon 
\cite{Lyubushkin:08a},
aluminium 
\cite{Belikov:81,Belikov:82b,Belikov:85a,Belikov:85b},
argon
\cite{Martinez:07},
iron
\cite{Suwonjandee:04},
steel
\cite{Kustom:69a},
propane
\cite{Budagov:69c},
freon
\cite{Young:67,Eichten:73a,Rollier:75,Bonetti:77,Makeev:81,Grabosch:88,Brunner:90,Ammosov:92},
and also propane--freon
\cite{Rollier:78,Pohl:79b,Armenise:79a} and
neon--hydrogen 
\cite{Asratyan:84a,Asratyan:84b,Ammosov:87b}
mixtures.
The recent MiniBooNE 2007 datapoint \cite{Aguilar-Arevalo:07} (carbon target) estimated from
the reported value of $M_A$ is also shown in
Fig.\ \ref{Fig:sQES_101.0.00.301.01_1_BBBA25_NT_PART1_PRD}
for comparison.

The compilation does not include obviously obsolete data
(e.g., ANL 1972 \cite{Mann:72}, CERN HLBC 1965/1966 \cite{Burmeister:65,Franzinetti:66}), 
as well as the data identical to those reported in the posterior publications of the same
experimental groups
(e.g., FNAL 1982 \cite{Asratyan:82}, GGM 1978 \cite{Dewit:78}, IHEP-ITEP 1983 \cite{Belikov:83},
IHEP SKAT 1986 \cite{Grabosch:86b}).
The early results of the NOMAD experiment reported in Refs.\ \cite{Petti:04,Lyubushkin:06},
have been considerably revised (mainly due to corrections in nuclear Monte Carlo) \cite{Lyubushkin:08a};
the datapoints shown in Fig.\ \ref{Fig:sQES_101.0.00.301.01_1_BBBA25_NT_PART1_PRD}
are still \emph{preliminary} and are reproduced here by permission of the NOMAD Collaboration. 

All the deuterium data quoted in Fig.\ \ref{Fig:sQES_101.0.00.301.01_1_BBBA25_NT_PART1_PRD}
and freon data in Fig.\ \ref{Fig:sQES_101.0.00.301.01_1_BBBA25_NT_PART2_PRD}
were converted to a free nucleon target by the experimenters.%
\footnote{The nuclear corrections applied to the deuterium data under consideration,
          were treated according to Singh \cite{Singh:72}.
          The nuclear effects for the freon data were modeled using a Fermi gas approach.
         }
The BNL 1981 experiment \cite{Baker:81b} had reported the $E_\nu$ and $Q^2$ dependencies
of $M_A$ extracted from a fit of the experimental $Q^2$ distribution rather than the cross
section; we quote the BNL 1981 cross section recalculated from $M_A$ by Kitagaki et al.\ \cite{Kitagaki:83}.
Similarly, the FNAL 1984 rectangle \cite{Asratyan:84a,Asratyan:84b} and FNAL 1987 datapoint \cite{Ammosov:87b}
were calculated by the experimenters (for free proton target) using the $M_A$ value extracted from the measured
$Q^2$ distribution of $\overline\nu_{\mu}$ events recorded in the Fermilab 15' bubble chamber filled with
a heavy neon-hydrogen mixture.
The data from several freon experiments (e.g., \cite{Young:67,Eichten:73a,Perkins:75}) reported in the
original papers in units cm$^2$ per nucleon of freon nucleus, were converted to the standard units.

All solid curves shown in the figures
were calculated using the BBBA(07) model for vector form factors with $d/u=0$
and always correspond to the best fit value
\begin{equation}\label{M_A_Global}
M_A^{\nu+\overline\nu} = 0.999 \pm 0.011~\text{GeV} \quad (\chi^2/\text{NDF} \approx 1.3)
\end{equation}
obtained from the global fit of neutrino and antineutrino data on the total and differential
cross sections (see Table \ref{table:M_A}).
We do not show the cross sections calculated with the GKex(05) model since the difference
will be practically invisible.

The dashed curves in Fig.\ \ref{Fig:sQES_101.0.00.301.01_1_BBBA25_NT_PART1_PRD}
are calculated with the $M_A$ values extracted from the best fit to the (preliminary) NOMAD
total cross section data alone \cite{Lyubushkin:08a}:
\begin{equation}
\label{M_A_NOMAD}
\begin{aligned}
M_A^{\nu}          & = 1.05 \pm 0.02_{\text{stat}} \pm 0.07_{\text{syst}}~\text{GeV},\\
M_A^{\overline\nu} & = 1.06 \pm 0.07_{\text{stat}} \pm 0.12_{\text{syst}}~\text{GeV},
\end{aligned}
\end{equation}
both agree with the global fit value \eqref{M_A_Global}.
Note that these results were obtained with the GKex(05) vector form factors.
Fitting the NOMAD data with the BBBA(07) form factors increases $M_A^{\nu}$ and $M_A^{\overline\nu}$
by about 0.8 and 0.9\%, respectively, that still remains well within the errors quoted in \eqref{M_A_NOMAD}.

As is seen from the figures, the obtained result, despite the non-optimal $\chi^2$ and
large spread of the data, is not in conflict with the main part of the data excluded
from the global fit.
Moreover, it well agrees with the world averaged value of
\begin{equation}\label{M_A_BBBA}
M_A = 1.014 \pm 0.014~\text{GeV},
\end{equation}
obtained in Ref.\ \cite{Bodek:07c} as a result of their reanalysis of the ``raw''
data from $\nu_{\mu}d$ and $\overline\nu_{\mu}H$ experiments
ANL 1973 \cite{Mann:73},
ANL 1977 \cite{S.Barish:77a},
ANL 1982 \cite{Miller:82},
BNL 1980 \cite{Fanourakis:80},
BNL 1981 \cite{Baker:81b},
BNL 1983 \cite{Baker:83a},
BNL 1990 \cite{Kitagaki:90},
FNAL 1983 \cite{Kitagaki:83},
CERN BEBC 1990 \cite{Allasia:90},
and from pion electroproduction experiments after corrections for hadronic effects.
Note that the values of $M_A$ re-extracted in Ref.\ \cite{Bodek:07c} from each $\nu_{\mu}d$
experiment separately spread between $0.97\pm0.05$ and $1.04\pm0.06$~GeV.
It exceeds the difference between the results of our analysis of data on total and differential
cross sections.
Both analyses use the same BBBA(07) model and mutually supplement each other, since they
practically do not overlap in the adopted data sets.
Formal averaging of the values \eqref{M_A_Global} and \eqref{M_A_BBBA} yields
\begin{equation*}\label{M_A_combined}
M_A = 1.006 \pm 0.009~\text{GeV}.
\end{equation*}

\subsection{Are $M_A^{\nu}$ and $M_A^{\overline\nu}$ really different?}

According to the global fit (see Table \ref{table:M_A}), the difference between the values of
$M_A^{\nu}$ and $M_A^{\overline\nu}$ obtained by fitting the neutrino and antineutrino data separately,
reaches about 3.5\% for BBBA(07) and about 3.2\% for GKex(05) that is above the statistical error
in determination of $M_A^{\nu}$ and $M_A^{\overline\nu}$.
However, taking into account the systematic difference between the fits of total and differential
cross section data, as well as high values of $\chi^2/\text{NDF}$, this difference cannot be
considered statistically significant. Furthermore, the fit to the antineutrino data is not
stable relative to including/excluding some data subsets.
In particular, as is seen from Fig.\ \ref{Fig:sQES_101.0.00.301.01_1_BBBA25_NT_PART1_PRD},
the total NuTeV cross sections per nucleon bound in iron, averaged over the
energy range $E_{\nu,\overline\nu}=30\div300~\text{GeV}$ 
\[
\begin{aligned}
\overline{\sigma}(\nu_{\mu}n\to\mu^-p)
& = \left(0.94 \pm 0.03_{\text{stat}} \pm 0.07_{\text{syst}}\right)\!\times\!10^{-38}\,\text{cm}^2, \\
\overline{\sigma}(\overline\nu_{\mu}p\to\mu^+n)
& = \left(1.12 \pm 0.04_{\text{stat}} \pm 0.10_{\text{syst}}\right)\!\times\!10^{-38}\,\text{cm}^2
\end{aligned}
\]
(shown in Fig.\ \ref{Fig:sQES_101.0.00.301.01_1_BBBA25_NT_PART1_PRD} by rectangles) notably exceed the
corresponding best fit curves whereby the NuTeV data \cite{Suwonjandee:04} strongly affects the global
fit values of $M_A^{\nu}$ and $M_A^{\overline\nu}$.

\begin{table*}[htb]
\centering
\caption{\label{table:M_A_Bodek}
         The same as in Table~\protect\ref{table:M_A} but after exclusion of the datasets from experiments
         with non-active targets (NuTeV~1984 \cite{Suwonjandee:04}, IHEP-ITEP~1981,82,85
         \cite{Belikov:81,Belikov:82b,Belikov:85b}) and the lowest-energy data of CERN 1967
         \cite{Young:67} (see text for details).
\hfill}
\begin{tabular}{ccc@{\qquad\qquad}ccc}
\hline\noalign{\smallskip}
\multicolumn{3}{c}{\bf BBBA(07)}                                  &  \multicolumn{3}{c}{\bf GKex(05)}                                 \\
\noalign{\smallskip}
\hline\noalign{\smallskip}
$M_A^\nu$       & $M_A^{\overline\nu}$ & $M_A^{\nu+\overline\nu}$ & $M_A^\nu$       & $M_A^{\overline\nu}$ & $M_A^{\nu,\overline\nu}$ \\
\noalign{\smallskip}
\hline\noalign{\smallskip}
\multicolumn{6}{l}{\bf Fit to the total cross sections:}                                                                              \\
\noalign{\smallskip}
$0.986\pm0.021$ & $0.855\pm0.046$      & $0.958\pm0.019$          & $0.977\pm0.021$ & $0.837\pm0.046$      & $0.948\pm0.019$          \\
$(42/52)$       & $(38/35)$            & $(88/88)$                & $(42/52)$       & $(38/35)$            & $(89/88)$                \\
\noalign{\smallskip}
\hline\noalign{\smallskip}
\multicolumn{6}{l}{\bf Fit to the differential cross sections:}                                                                       \\
\noalign{\smallskip}
\noalign{\smallskip}
$0.966\pm0.024$ & $0.971\pm0.042$      & $0.967\pm0.021$          & $0.963\pm0.024$ & $0.959\pm0.043$      & $0.962\pm0.021$          \\
$(33/33)$       & $(16/22)$            & $(49/56)$                & $(34/33)$       & $(15/22)$            & $(49/56)$                \\
\hline\noalign{\smallskip}
\multicolumn{6}{l}{\bf Fit to the total and differential cross sections:}                                                             \\
\noalign{\smallskip}
$0.977\pm0.016$ & $0.912\pm0.030$      & $0.962\pm0.014$          & $0.971\pm0.016$ & $0.896\pm0.031$      & $0.954\pm0.014$          \\
$(75/86)$       & $(58/58)$            & $(137/145)$              & $(76/86)$       & $(57/58)$            & $(138/145)$              \\
\noalign{\smallskip}
\hline\noalign{\smallskip}
\end{tabular}
\end{table*}

To clarify this point further, we have performed additional fits, in which the datasets obtained in experiments
with non-active targets have been removed. Namely, we excluded the highest energy NuTeV total cross section
data (iron target) \cite{Suwonjandee:04} and the data on differential cross sections measured with the
IHEP-ITEP spark chamber detector with aluminium filters \cite{Belikov:81,Belikov:82b,Belikov:85b},
since these experiments do not have an active target to measure recoil hadrons and surely remove resonance
background. In order to minimize possible uncertainties in nuclear corrections, the lowest-energy CERN 1967
total cross section data (freon target) \cite{Young:67} were also excluded from these fits.
The results of this analysis are summarized in Table \ref{table:M_A_Bodek}.
It is seen that the additional reduction of the dataset essentially \emph{decreases} the resulting values of $M_A$.
Concurrently it improves the statistical quality of the fits to the total cross section data, while slightly
increases the $\chi^2/\text{NDF}$ for the fit to the differential cross sections.
Besides that, the $M_A$ values extracted from the total and differential cross sections become bit more consistent.
The differences between $M_A^{\nu}$ and $M_A^{\overline\nu}$
[-65~MeV for BBBA(07) and -75~MeV for GKex(05)]
become \emph{opposite in sign} to those obtained from our ``default'' fit performed with the full dataset.
However, both $M_A^{\nu}$ and $M_A^{\overline\nu}$ values are still compatible, within the $1\sigma$
deviation, with the average value of $M_A^{\nu+\overline\nu}$.
So we may reckon that
\begin{itemize}
\item[ (i)] the axial mass extraction is rather responsive to the choice of the data subsets and
\item[(ii)] the current experimental data cannot definitely confirm or disconfirm possible difference between
            the axial masses extracted from experiments with neutrino and antineutrino beams.
\end{itemize}
Similar fit performed for the differential cross section data only, from which all the $\nu_{\mu}d$ data
were excluded, leads to an \emph{increase} of $M_A^{\nu}$ by about 4.2\% (4.4\%) for BBBA(07) (GKex(05)).
However, the statistical error of this fit increases too. Including into this fit the non-deuterium data
on total cross sections diminish the increase of $M_A^{\nu}$ to about 1.2\% for both BBBA(07) and GKex(05).
Hence, the above conclusions remain essentially unchanged. 

\subsection{Further details on differential cross section data}

As is known from the comparison with the low-energy electron-nucleus scattering data,
the RFG description of the low-$Q^2$ region is not enough accurate especially at energies below
$\sim2~\text{GeV}$ (for recent discussion, see, e.g., Refs.\ \cite{Butkevich:05,Butkevich:07} and references therein).
Moreover, the shape of $d\sigma/dQ^2$ at $Q^2\lesssim0.1~\text{GeV}^2$ is slowly sensitive to
variations of $M_A$ (see below).
Thus, in order to minimize possible uncertainties due to nuclear effects, the points with
$Q^2<0.15~\text{GeV}^2$ were rejected from the fit of the differential cross section dataset.
Leaving these points in the fit would lead to a \emph{decrease} of the output values of $M_A^{\nu}$,
$M_A^{\overline\nu}$, and $M_A^{\nu,\overline\nu}$ obtained from the $d\sigma/dQ^2$ dataset
by, respectively, 1.8, 3.3, and 2.2\% for BBBA(07) and 2.0, 4.0, and 2.6\% for GKex(05)
form factors.
The corresponding decrease of $M_A$ derived from the full dataset ($\sigma$ and $d\sigma/dQ^2$)
is clearly less essential: respectively, 0.7, 1.3, and 0.9\% for BBBA(07) and
0.7, 1.5, and 1.0\% for GKex(05).

Of course, the mentioned uncertainty still remains in the RFG calculations of the total
cross sections, since the contribution from the low-$Q^2$ region is essential at low energies.
To illustrate this, we show in Fig.\ \ref{Fig:RQES(Q2)_BBBA25_C} the relative contribution of
the region $Q^2<Q_1^2$ into the total cross section, $R\left(Q_1^2\right)=\sigma\left(Q^2<Q_1^2\right)/\sigma$,
as a function of $Q_1^2$, evaluated for $\nu_\mu$ and $\overline\nu_\nu$ QE interactions
with carbon at several (anti)neutrino energies using $M_A=1~\text{GeV}$.%
\footnote{Here $\sigma\left(Q^2<Q_1^2\right)$ is defined as an integral of $d\sigma/dQ^2$
          from the kinematical minimum of $Q^2$ to $Q^2=Q_1^2$.
         }
It is seen that for neutrino-nucleus interactions $R\lesssim0.25$ as $Q_1^2<0.15~\text{GeV}^2$ and
$E_\nu>0.7~\text{GeV}$ that is for all energies of our current interest.
As a result, a few percent error expected in $d\sigma/dQ^2$ due to inaccuracy of the RFG model for the low-$Q^2$
region, becomes nearly negligible in the total cross section.
However it is not the case for antineutrino interactions, for which the ratio $R\left(Q_1^2=0.15~\text{GeV}^2\right)$
becomes reasonably small ($R\lesssim0.3$) only for $E_\nu\gtrsim2~\text{GeV}$.
Therefore the lower energy antineutrino total cross section data may bias an uncontrolled (while still small)
additional uncertainty.
Fortunately, the major part of the data participated in the global fit satisfies the above conditions and our
examination demonstrates that the related uncertainty is not weighty.

\begin{figure*}[b]
\centering
\includegraphics[width=0.48\linewidth]{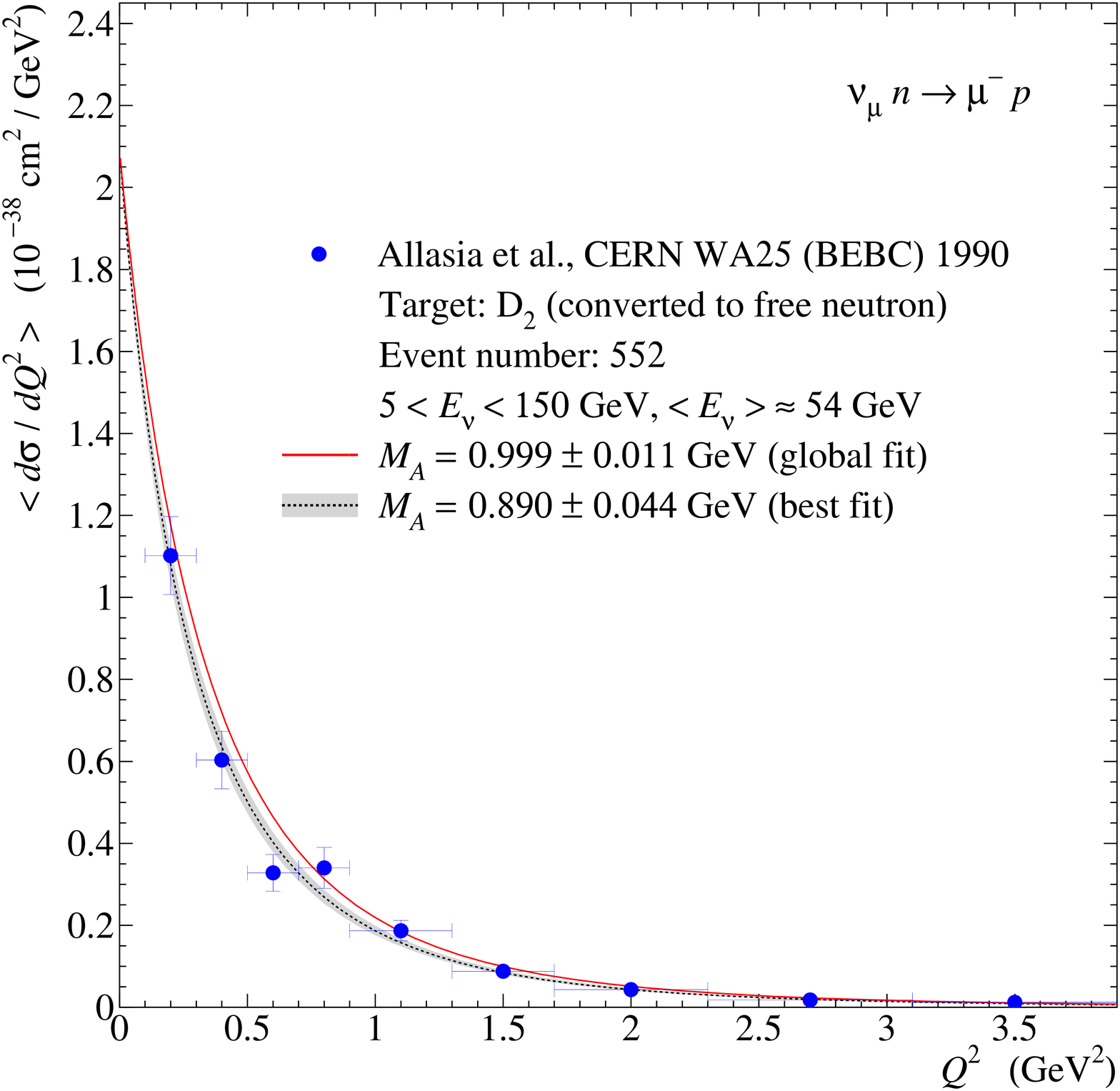}
\caption{Flux-weighted differential cross section for $\nu_{\mu}n\to\mu^-p$ 
         measured in the WA25 experiment with the CERN bubble chamber BEBC
         filled with deuterium and exposed to high-energy $\nu_{\mu}$ beam at
         the CERN-SPS \cite{Allasia:90}.
         The data were converted to a free neutron target by the authors of the
         experiment.
         The curves are the calculated cross sections averaged over the
         experimental $\nu_{\mu}$ energy spectrum borrowed from Ref.\ \cite{Allasia:84}.
         The energy range and estimated mean energy are given in the legend.
         The dashed curves are for the best fit to the WA25 data, while
         the solid curves correspond to the global fit to all QES data.
         Shaded band represents $1\sigma$ deviation from the best-fitted value
         of $M_A$ given in the legend.
        }
\label{Fig:dsQES_dQ2_Allasia_BEBC90}
\end{figure*}

\begin{figure*}[b]
\centering
\includegraphics[width=0.91\linewidth]{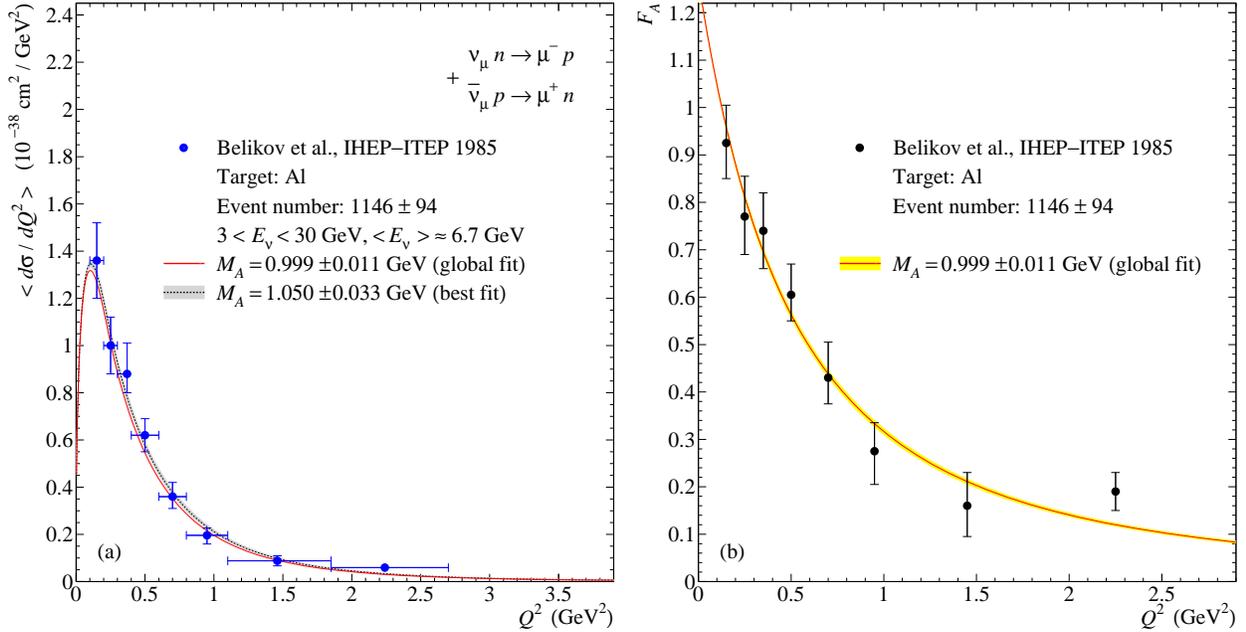}
\caption{Flux-weighted semisum of differential cross sections for
         $\nu_{\mu}n\to\mu^-p$ and $\overline\nu_{\mu}p\to\mu^+n$ (a) and
         axial-vector form factor $F_A(Q^2)$ (b) measured in the IHEP-ITEP experiment with
         a spark chamber detector with aluminium filters and exposed to the U70 broad-band
         $\nu_{\mu}$  and $\overline\nu_{\mu}$ beams of the Serpukhov PS \cite{Belikov:85b}.
         The error bars represent the total errors which include the overall systematic
         error of about 10\% (due mainly to uncertainties of the flux normalization
         and scanning/triggering efficiencies).
         The curves in panel (a) are the calculated semisum of the cross sections each
         averaged over the experimental $\nu_{\mu}$ and $\overline\nu_{\mu}$ energy
         spectra borrowed from Refs.\ \cite{Belikov:85b,Ammosov:92}.
         The energy range and estimated mean energy are given in the legend.
         The dashed curve is for the best fit to the quoted IHEP-ITEP data,
         while the solid curve corresponds to the global fit to all QES data.
         Shaded bands in panels (a) and (b) represent $1\sigma$ deviations from
         the best-fitted values of $M_A$ given in the legends.
        }
\label{Fig:dsQES_dQ2_Belikov_IHEP-ITEP85}
\end{figure*}

\begin{figure*}[t]
\centering
\includegraphics[width=0.91\linewidth]{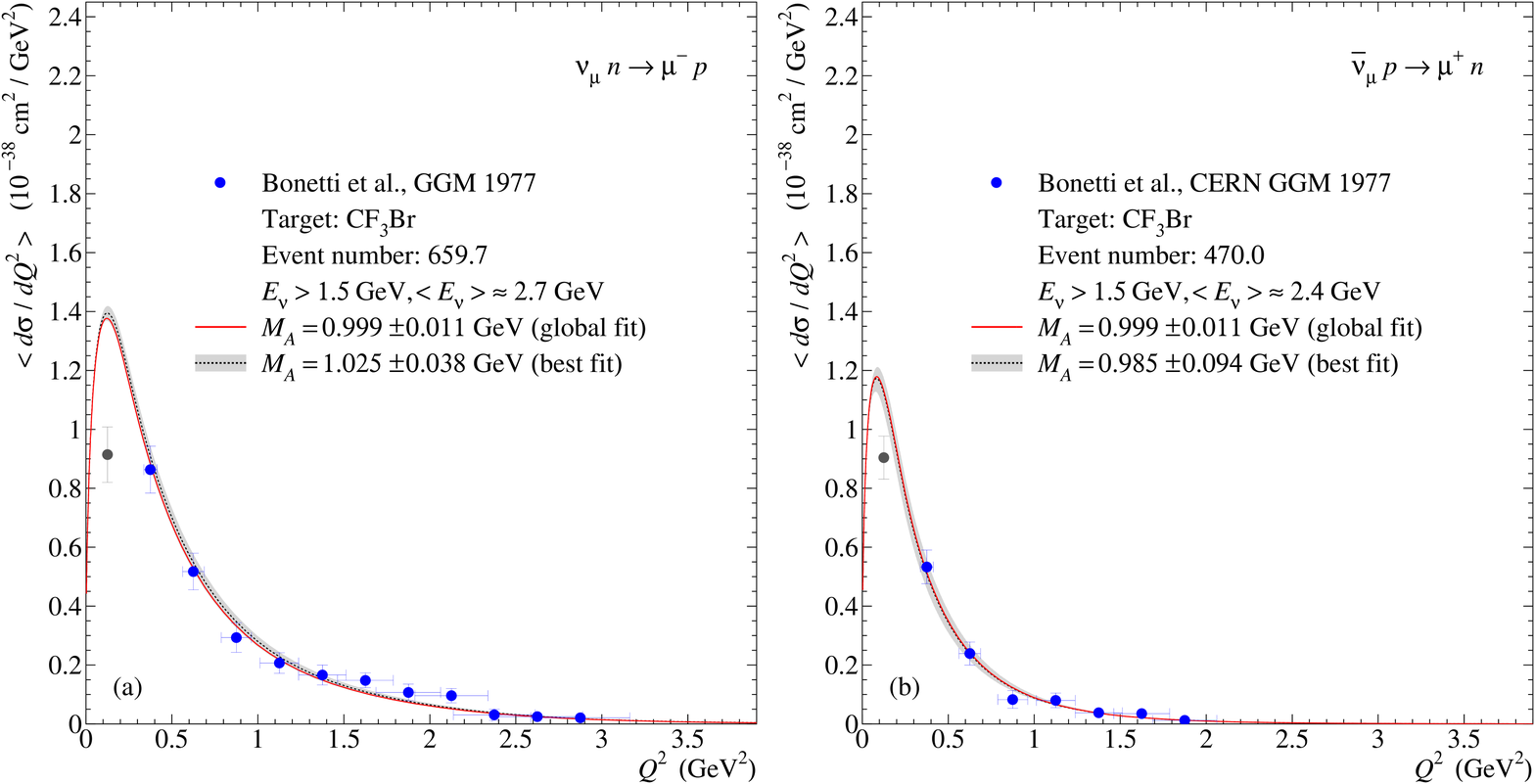}
\caption{Flux-weighted differential cross sections for
         $\nu_{\mu}n\to\mu^-p$ (a) and $\overline\nu_{\mu}p\to\mu^+n$ (b)
         measured with the heavy-liquid bubble chamber Gargamelle filled with
         heavy freon and exposed to the CERN-PS $\nu_{\mu}$ and 
         $\overline\nu_{\mu}$ beams \cite{Bonetti:77,Musset:78}.
         The error bars contain the statistical fluctuation and the
         indetermination on the $\nu_{\mu}$ and $\overline\nu_{\mu}$ fluxes.
         The curves are the calculated cross sections averaged over the
         experimental $\nu_{\mu}$ and $\overline\nu_{\mu}$ energy spectra
         given in Ref.\ \cite{Bonetti:77}.
         Only the events with $E_{\nu,\overline{\nu}}>1.5$~GeV were accepted.
         The dashed curves are for the best fit to the GGM~1977 data, while
         the solid curves correspond to the global fit to all QES data.
         The points shown by grey symbols are excluded from the fits (see text).
         Shaded bands represent $1\sigma$ deviations from the best-fitted values
         of $M_A$ given in the legends.
        }
\label{Fig:dsQES_dQ2_Bonetti_GGM77}
\end{figure*}

\begin{figure*}[b]
\centering
\includegraphics[width=0.91\linewidth]{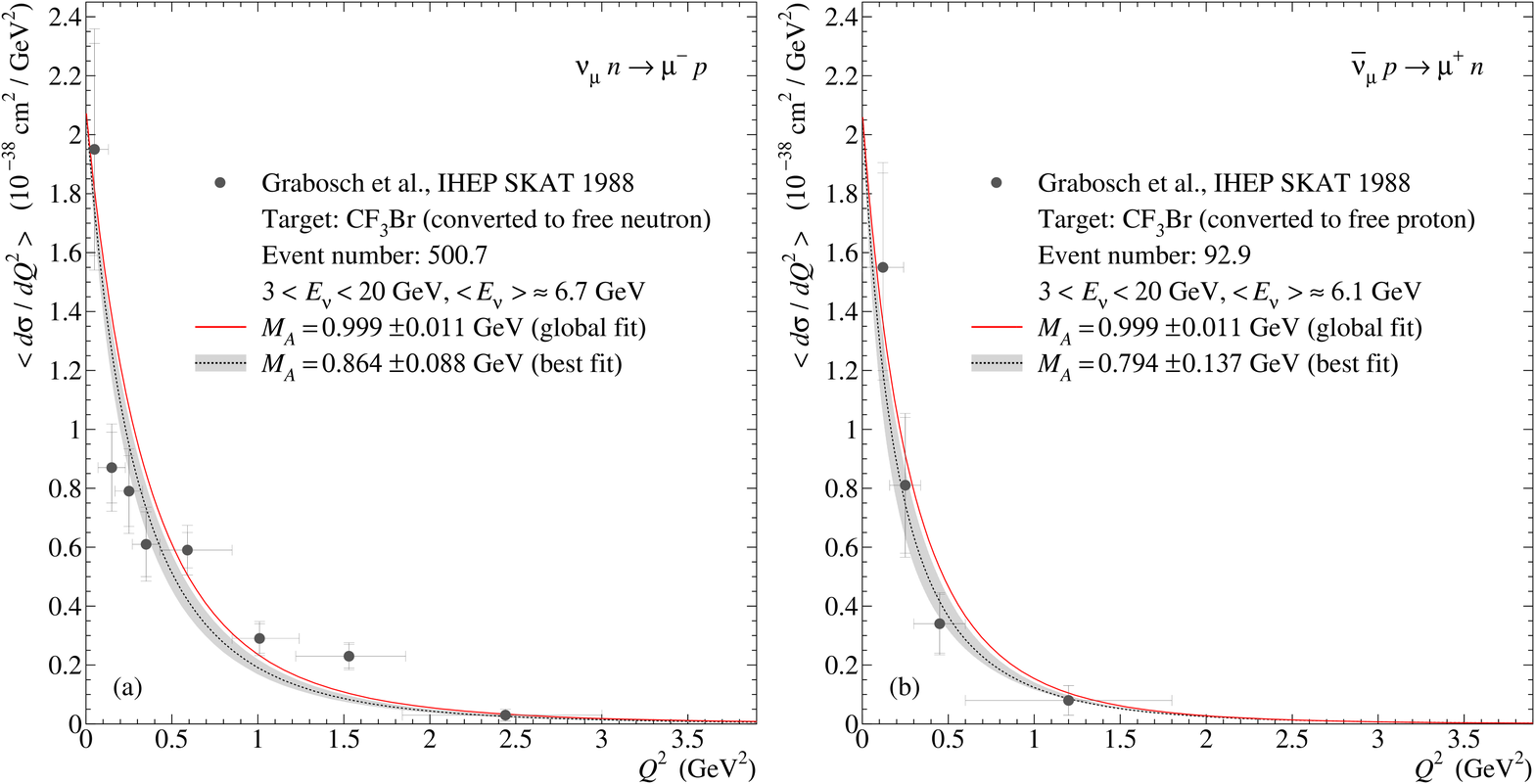}
\caption{Flux-weighted differential cross sections for
         $\nu_{\mu}n\to\mu^-p$ (a) and $\overline\nu_{\mu}p\to\mu^+n$ (b)
         measured with the freon filled bubble chamber SKAT exposed to the U70
         broad-band $\nu_{\mu}$ and $\overline\nu_{\mu}$ beams of the Serpukhov
         PS \cite{Grabosch:88,Ammosov:92}
         (see also Refs.\ \cite{Grabosch:86b} for the earlier analyses
         of the same data sample).
         The data were converted to a free nucleon target by the authors of the
         experiment.
         The inner and outer bars indicate statistical and total errors,
         respectively; the systematic error includes the uncertainties due to
         the cross section normalization and nuclear Monte Carlo.
         The curves are the calculated cross sections averaged over the
         experimental $\nu_{\mu}$ and $\overline\nu_{\mu}$ energy spectra
         borrowed from Ref.\ \cite{Ammosov:92}.
         The energy range and estimated mean energies are given in the legends.
         The dashed curves are for the best fit to the SKAT~1988 data, while
         the solid curves correspond to the global fit to all QES data
         (the SKAT~1988 data are excluded from the global fit).
         Shaded bands represent $1\sigma$ deviations from the best-fitted values of
         $M_A$ given in the legends.
        }
\label{Fig:dsQES_dQ2_Grabosch_SKAT88}
\end{figure*}

\begin{figure*}[t]
\centering
\includegraphics[width=0.91\linewidth]{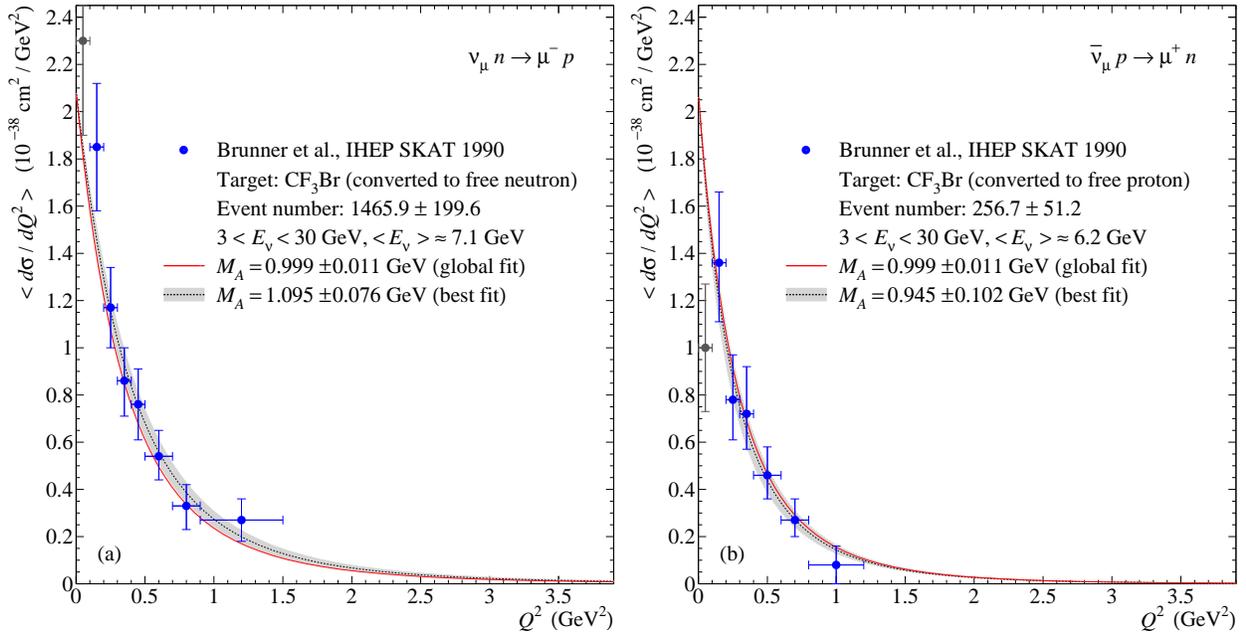}
\caption{Flux-weighted differential cross sections for
         $\nu_{\mu}n\to\mu^-p$ (a) and $\overline\nu_{\mu}p\to\mu^+n$ (b)
         measured with the freon filled bubble chamber SKAT exposed to the U70
         broad-band $\nu_{\mu}$ and $\overline\nu_{\mu}$ beams of the Serpukhov
         PS \cite{Brunner:90}.
         The data were converted to a free nucleon target by the authors of the
         experiment.
         The inner and outer bars indicate statistical and total errors,
         respectively; the systematic error includes the uncertainties due to
         the cross section normalization and nuclear Monte Carlo.
         The curves are the calculated cross sections averaged over the
         experimental $\nu_{\mu}$ and $\overline\nu_{\mu}$ energy spectra
         borrowed from Ref.\ \cite{Ammosov:92}.
         The energy range and estimated mean energies are given in the legends.
         The dashed curves are for the best fit to the SKAT~1990 data, while
         the solid curves correspond to the global fit to all QES data.
         The points shown by grey symbols are excluded from the fits (see text).
         Shaded bands represent $1\sigma$ deviations from the best-fitted values
         of $M_A$ given in the legends.
        }
\label{Fig:dsQES_dQ2_Brunner_SKAT90}
\end{figure*}

\begin{figure*}[b]
\centering
\includegraphics[width=0.91\linewidth]{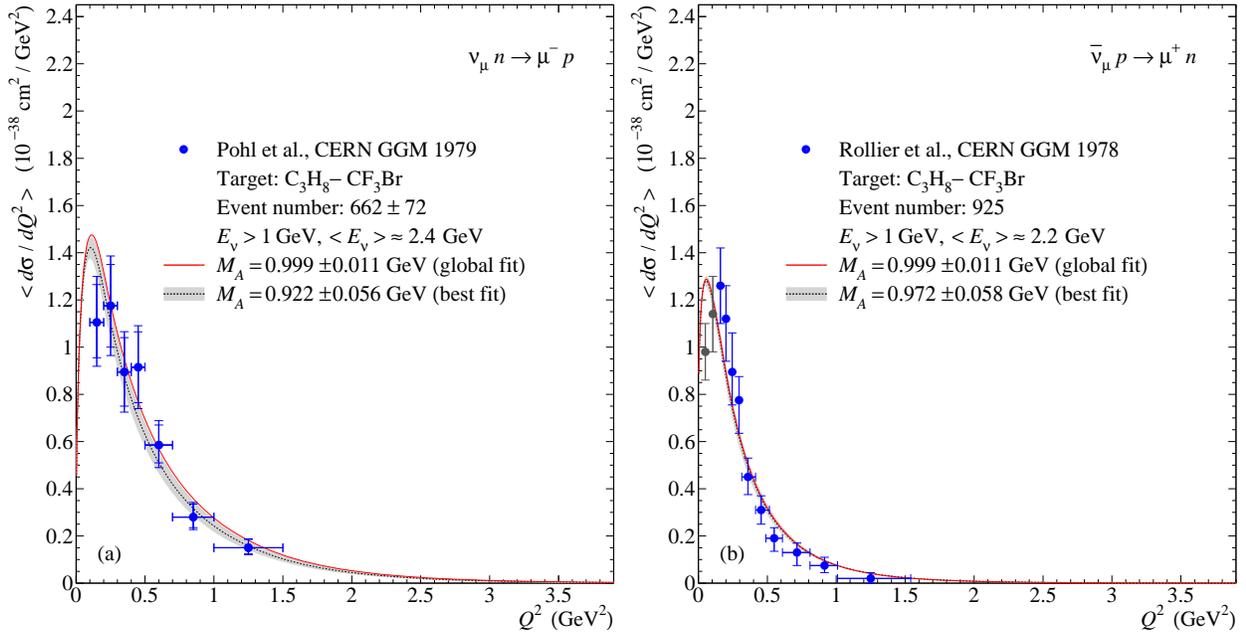}
\caption{Flux-weighted differential cross sections for
         $\nu_{\mu}n\to\mu^-p$ (a) and $\overline\nu_{\mu}p\to\mu^+n$ (b)
         measured with the bubble chamber Gargamelle filled with light
         propane--freon mixture and exposed to the CERN-PS $\nu_{\mu}$ and 
         $\overline\nu_{\mu}$ beams \cite{Pohl:79b,Rollier:78}.
         The inner and outer bars in panel (a) indicate statistical and total errors,
         respectively; the error bars in panel (b) contain the statistical fluctuation
         and the indetermination on the $\overline\nu_{\mu}$ flux.
         The curves are the calculated cross sections averaged over the
         experimental $\nu_{\mu}$ and $\overline\nu_{\mu}$ energy spectra
         given in Refs.\ \cite{Bonetti:77} and \cite{Armenise:79a}, respectively.
         Only the events with $E_{\nu,\overline{\nu}}>1$~GeV were accepted.
         The dashed curves are for the best fit to the GGM data, while
         the solid curves correspond to the global fit to all QES data.
         The points shown by grey symbols are excluded from the fits (see text).
         Shaded bands represent $1\sigma$ deviations from the best-fitted values
         of $M_A$ given in the legends.
        }
\label{Fig:dsQES_dQ2_Pohl_GGM79_Rollier_GGM78}
\end{figure*}

Figures \ref{Fig:dsQES_dQ2_Allasia_BEBC90}--\ref{Fig:dsQES_dQ2_Belikov_IHEP-ITEP85}\,(a) and
\ref{Fig:dsQES_dQ2_Bonetti_GGM77}--\ref{Fig:dsQES_dQ2_Pohl_GGM79_Rollier_GGM78}
represent the spectrum-ave\-ra\-ged differential cross sections for several nuclear targets:
deuterium
(Fig.\ \ref{Fig:dsQES_dQ2_Allasia_BEBC90})
\cite{Allasia:90},
aluminium
(Figs.\ \ref{Fig:dsQES_dQ2_Belikov_IHEP-ITEP81,82} and \ref{Fig:dsQES_dQ2_Belikov_IHEP-ITEP85}\,(a))
\cite{Belikov:81,Belikov:82b,Belikov:85b},
freon    
(Figs.\ \ref{Fig:dsQES_dQ2_Bonetti_GGM77}, \ref{Fig:dsQES_dQ2_Grabosch_SKAT88} and
\ref{Fig:dsQES_dQ2_Brunner_SKAT90})
\cite{Bonetti:77,Grabosch:88,Brunner:90,Musset:78,Ammosov:92}, and
propane-freon mixture
(Fig.\ \ref{Fig:dsQES_dQ2_Pohl_GGM79_Rollier_GGM78})
\cite{Pohl:79b,Rollier:78}.
In Fig.~\ref{Fig:dsQES_dQ2_Belikov_IHEP-ITEP85}\,(b) we show (for illustrative purposes only)
the axial-vector form factor extracted in the IHEP-ITEP spark chamber experiment \cite{Belikov:85b}.
All the quoted data, except those from Ref.\ \cite{Grabosch:88} (superseded by the data from
the more recent publication by the SKAT Collaboration~\cite{Brunner:90}), model-dependent
IHEP-ITEP data on $F_A(Q^2)$~\cite{Belikov:85b}, and a few rejected low-$Q^2$ datapoints,
participate in the global fit.
We show the cross sections calculated with $M_A$ obtained by individual fits to the data of each
experiment alone and compare these against the cross sections evaluated with the global-fit value
of $M_A$. All the details are recounted in the captions and legends of the figures.
The comparison demonstrates that the individual and global fits generally do not contradict
each other. The differences are within the experimental errors and are not of systematic nature.

\begin{figure*}[t]
\centering
\includegraphics[width=\linewidth]{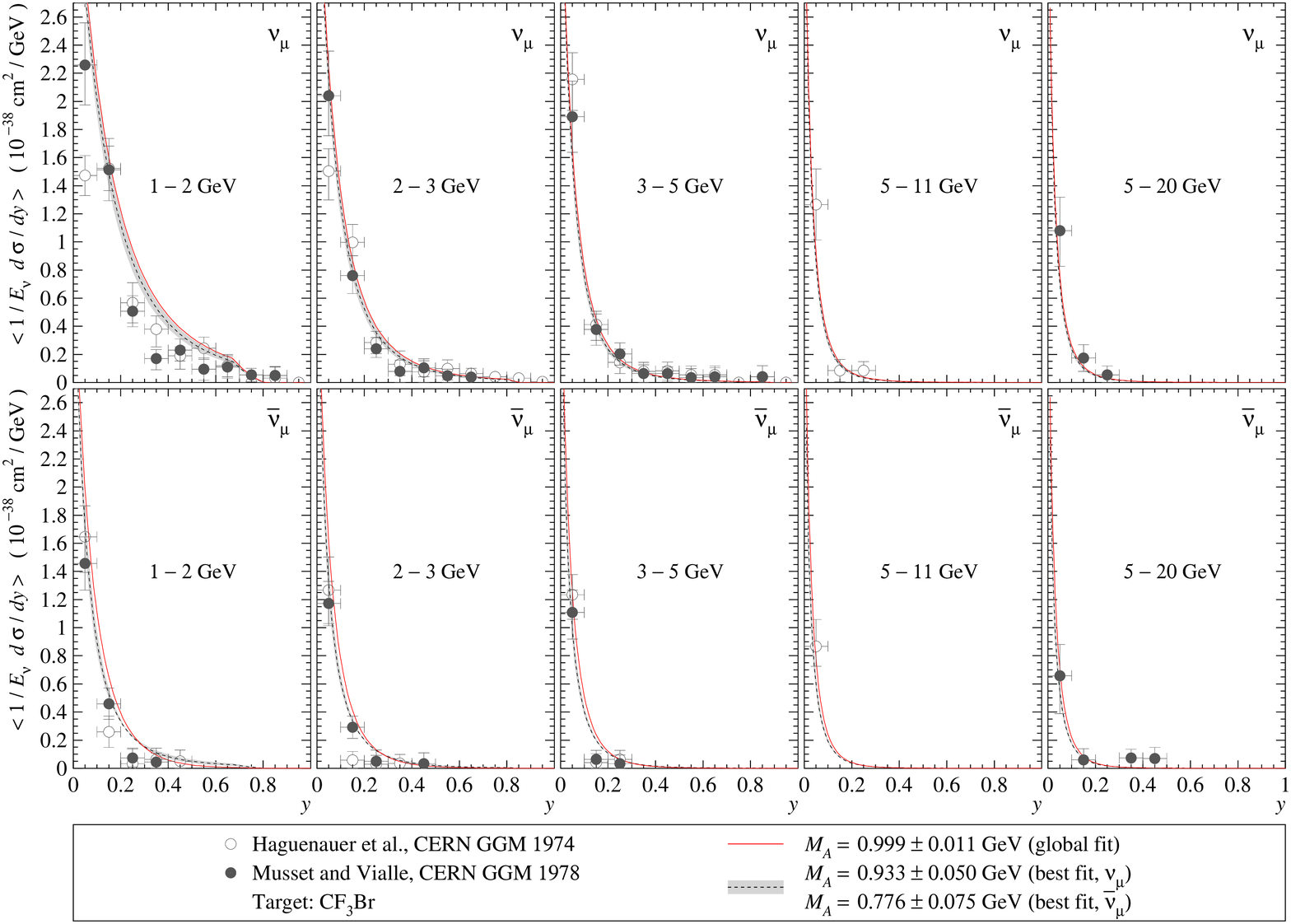}
\caption{Flux-weighted differential cross sections
         $(1/E_\nu)d\sigma(\nu_{\mu}n\to\mu^-p)/dy$ (top panels) and
         $(1/E_\nu)d\sigma(\overline\nu_{\mu}p\to\mu^+n)/dy$ (bottom panels)
         measured with the heavy freon filled bubble chamber Gargamelle exposed
         to the wide-band CERN-PS $\nu_{\mu}$ and $\overline\nu_{\mu}$ beams
         \cite{Haguenauer:74,Musset:78}.
         The data from Refs.~\cite{Haguenauer:74} (range $5-11~\text{GeV}$) and
         \cite{Musset:78} (ranges $1-2$, $2-3$, $3-5$, and $5-20~\text{GeV}$)
         represent two different analyses of the same data sample
         (see also Refs.\ \cite{Sciulli:74a,Deden:75,Baltay:94} for other versions).
         The measured cross sections were converted to a free nucleon target
         by the authors of the experiment.
         The quoted error bars are the total errors which include the
         uncertainties in the $\nu_{\mu}$ and $\overline\nu_{\mu}$ fluxes
         and nuclear Monte Carlo.
         The curves are for the calculated cross sections averaged (for each energy
         range indicated in the panels) over the experimental $\nu_{\mu}$ and
         $\overline\nu_{\mu}$ energy spectra taken from Ref.\ \cite{Musset:78}.
         The dashed curves correspond to the $M_A$ values obtained by fitting
         the GGM~1978 data from the energy ranges $2-3$, $3-5$, and $5-20~\text{GeV}$
         and GGM~1974 data from the range $5-11~\text{GeV}$ (separately for neutrino
         and antineutrino cross sections).
         The range $1-2~\text{GeV}$ is excluded from the analysis in order to
         minimize the error in modelling the nuclear effects.
         The solid curves correspond to the global fit to all QES data
         (the GGM data are not included in this fit).
         Shaded bands represent $1\sigma$ deviations from the best-fitted
         values of $M_A$ given in the legend.
        }
\label{Fig:dsQES_dy_1_BBBA25_PRD}
\end{figure*}

As a further test of the global fit, we show in Fig.\ \ref{Fig:dsQES_dy_1_BBBA25_PRD}
the flux-weighted differential cross sections $d\sigma(\nu_{\mu}n\to\mu^-p)/dy$
and $d\sigma(\overline\nu_{\mu}p\to\mu^+n)/dy$ (divided by energy), which were measured with
the Gargamelle bubble chamber filled with liquid freon and exposed to the wide-band CERN-PS
$\nu_{\mu}$ and $\overline\nu_{\mu}$ beams. Several analyses of these data samples are
available from the literature (see Refs.\ \cite{Sciulli:74a,Haguenauer:74,Deden:75,Musset:78}
and also Ref.\ \cite{Baltay:94} for a review).
Figure \ref{Fig:dsQES_dy_1_BBBA25_PRD} shows two representative versions taken from
Refs.\ \cite{Haguenauer:74} and \cite{Musset:78} -- the preliminary and final results
of the GGM experiment, respectively. The data are shown for the five narrow instrumental ranges:
$1-2$, $2-3$, $3-5$, $5-11$, and $5-20$~GeV.
The measured cross sections were converted from freon to a free nucleon target by the experimenters,
after accounting for Fermi motion of the nucleons and Pauli suppression of quasielastic events.

For a qualitative comparison, we have performed individual fits to the GGM data,
separately for neutrino and antineutrino differential cross sections.
In order to reduce possible error introduced by RFG calculations of nuclear effects,
the energy range of $1-2~\text{GeV}$ has been excluded from this likelihood analysis.
As is seen from the figure, the $M_A$ value extracted from the neutrino subsample does not
contradict to that from the global fit, while it is not so for the antineutrino data subsample
where the discrepancy is essential.
This discrepancy can be attributed (at least, partially) to the vagueness of the 
model for nuclear effects used in the analyses of the GGM data.
Since the details of the GGM nuclear Monte Carlo are not available, we do not include this
data sample into the global fit.
We note, however, that the inclusion of these data (also without the low-energy datapoints)
into the fit only leads to a small \emph{decrease} of the output values of $M_A^{\nu}$,
$M_A^{\overline\nu}$, and $M_A^{\nu,\overline\nu}$ -- by, respectively, 0.4, 2.2, and 0.9\%
for BBBA(07) and 0.3, 2.0, and 0.8\% for GKex(05) form factors.
The corresponding $\chi^2/\text{NDF}$ values remain nearly the same.

\subsection{$Q^2$ distributions}

An additional fruitful set of available data is the $Q^2$ distributions $dN/dQ^2$ of the QES events
measured in several experiments with different nuclear targets. Usually just $dN/dQ^2$ is considered
as the observable most appropriate for extracting axial mass value, since it is less dependent
of the flux and spectrum uncertainties in comparison with the differential or total cross sections.
However, in comparison with the differential cross section, the $Q^2$ distribution has
two drawbacks: it contains an uncertainty due to normalization, and it is generally less
responsive to variations of $M_A$ at high $Q^2$. Figure \ref{Fig:Ratios} illustrates the second point.
It shows the $Q^2$ distributions and differential cross sections for $\nu_{\mu}$ and $\overline\nu_{\mu}$
quasielastic scattering off free nucleons, evaluated with different values of $M_A$ and normalized
to the corresponding quantities calculated with $M_A=1~\text{GeV}$.
The calculations are done with the fixed values of energy corresponding to the mean (anti)neutrino
beam energies in experiments \cite{Asratyan:84b,Budagov:69c,Armenise:79a,Makeev:81}.
It is seen from the figure that the region $Q^2\lesssim0.15~\text{GeV}^2$ strongly affected
by the nuclear effects, is sensitive to $M_A$ for $dN/dQ^2$ and less sensitive for $d\sigma/dQ^2$;
the situation is opposite for the high $Q^2$ region for which the nuclear corrections are less important.

\begin{figure*}[t]
\centering
\includegraphics[width=0.51\linewidth]{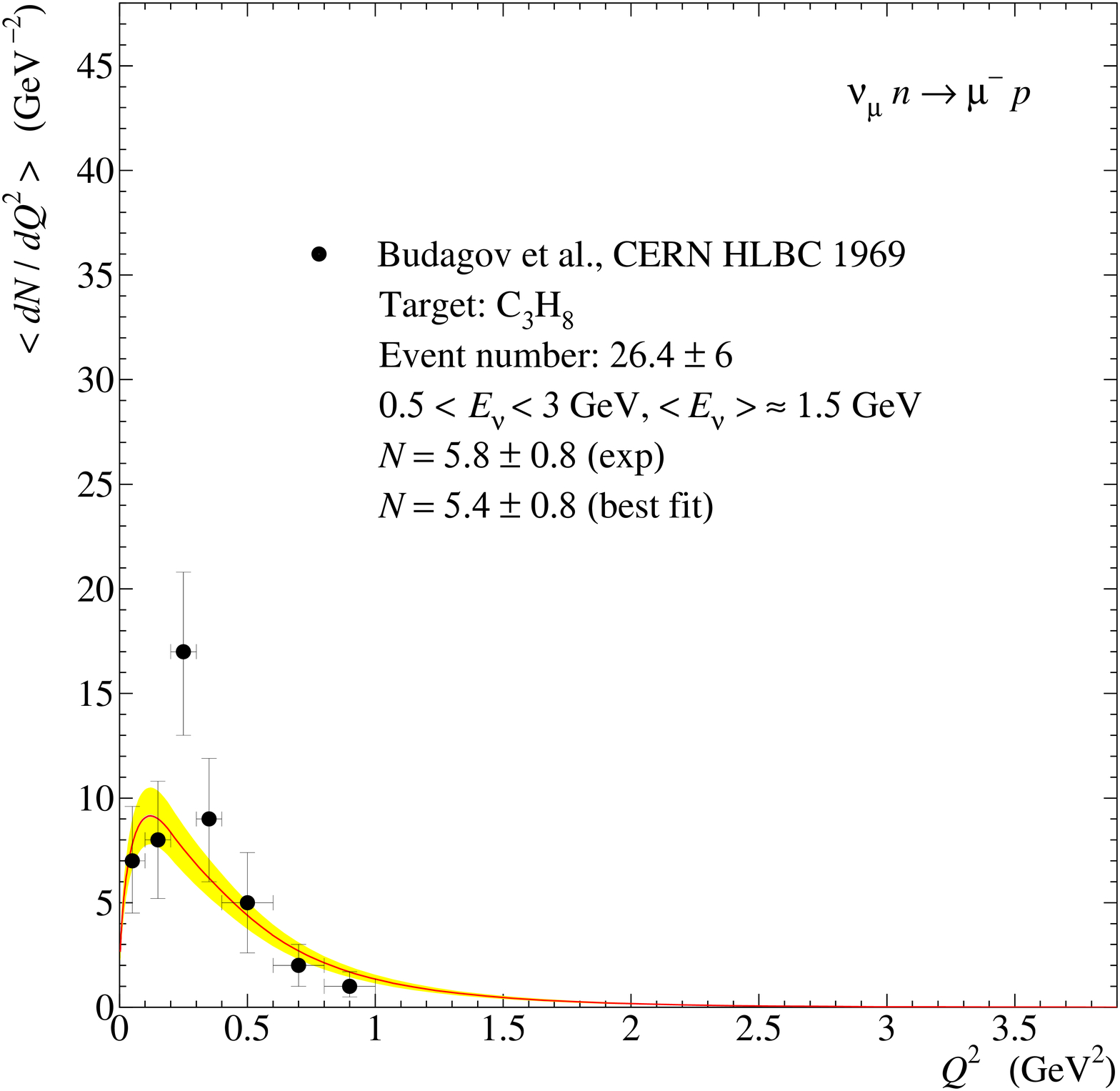}
\caption{Flux-weighted $Q^2$ distribution for $\nu_{\mu}n\to\mu^-p$
         measured with the CERN heavy-liquid bubble chamber (HLBC) filled with
         propane and exposed to the CERN PS $\nu_{\mu}$ beam \cite{Budagov:69c}.
         The curve is the distribution calculated with $M_A$ obtained from the
         global fit, averaged over the experimental $\nu_\mu$ energy spectrum
         from Ref.\ \cite{Budagov:69a}, and normalized to the HLBC~1969 data.
         The spectrum is estimated to be accurate within $\pm15\%$ (the error
         includes an estimate of systematic effects). 
         The energy range and estimated mean energy are given in the legends.
         Shaded band represents $1\sigma$ variation from the average due to
         uncertainties in $M_A$ and normalization factor $N$.
       }
\label{Fig:dNQES_dQ2_Budagov_HLBC69_1}
\vspace*{5mm}
\centering
\includegraphics[width=0.51\linewidth]{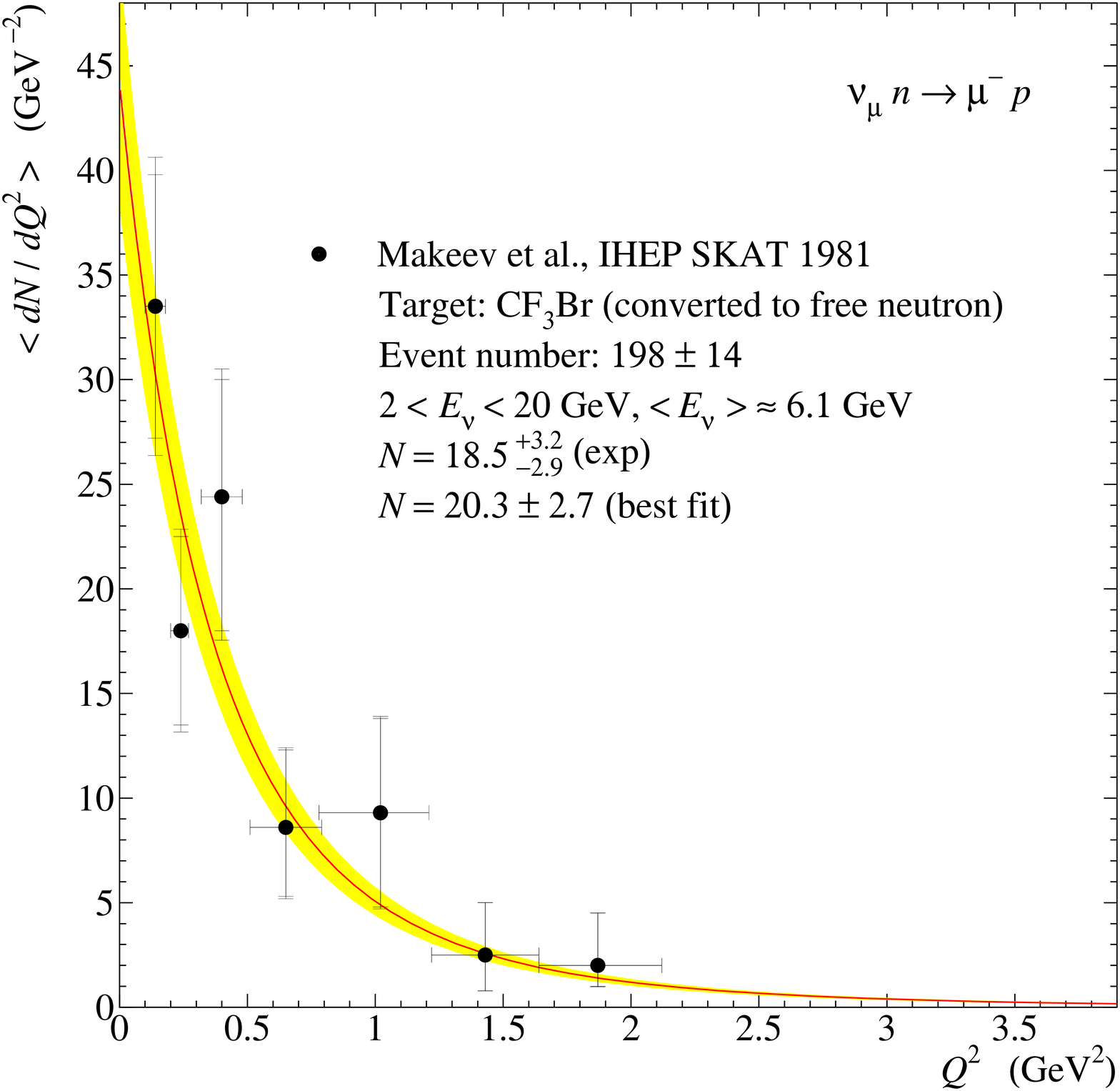}
\caption{Flux-weighted $Q^2$ distribution for $\nu_{\mu}n\to\mu^-p$
         measured with the freon filled bubble chamber SKAT exposed to the U70
         broad-band $\nu_{\mu}$ beam of the Serpukhov PS \cite{Makeev:81}.
         The data were converted to a free nucleon target by the authors of the
         experiment.
         The inner and outer bars indicate statistical and total errors,
         respectively; the systematic error includes the uncertainties due to
         the flux normalization and nuclear Monte Carlo.
         The curve is the distribution calculated with $M_A$ obtained from the
         global fit, averaged over the experimental $\nu_{\mu}$ energy spectrum
         from Ref.\ \cite{Ammosov:92}, and normalized to the SKAT~1981 data.
         The energy range and estimated mean energy are given in the legends.
         Shaded band represents $1\sigma$ variation from the average due to
         uncertainties in $M_A$ and normalization factor $N$.
       }
\label{Fig:dNQES_dQ2_Makeev_SKAT81_1}
\end{figure*}

\begin{figure*}[t]
\centering
\includegraphics[width=0.51\linewidth]{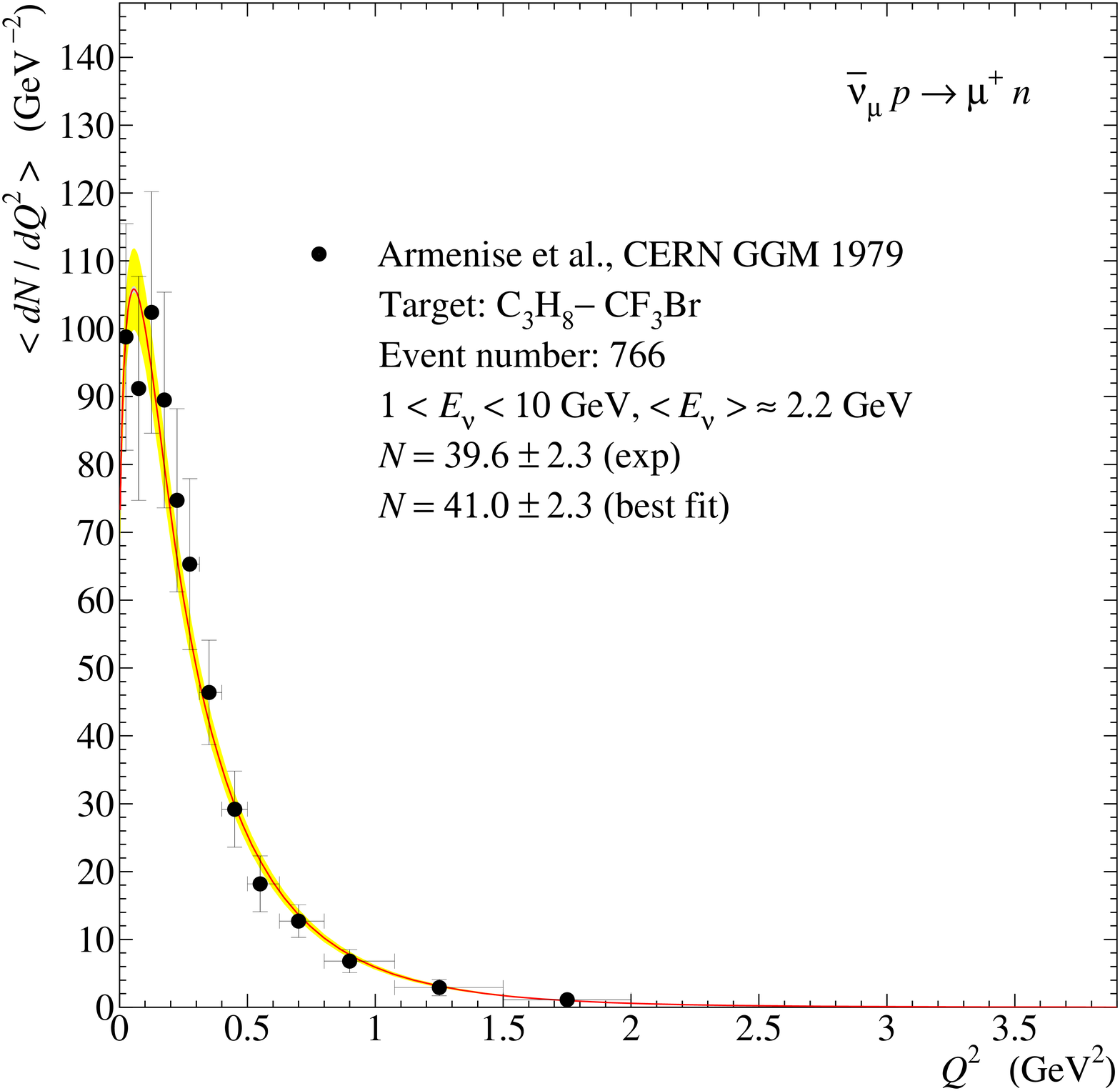}
\caption{Flux-weighted $Q^2$ distribution for $\overline\nu_{\mu}p\to\mu^+n$
         measured with the bubble chamber Gargamelle filled with light
         propane--freon mixture (87 mole per cent of propane) and exposed to
         the CERN-PS $\overline\nu_{\mu}$ beam \cite{Armenise:79a}.
         The error bars contain both statistical and systematic errors.
         The curve is the distribution calculated with $M_A$ obtained from the
         global fit, averaged over the experimental $\overline\nu_{\mu}$ energy
         spectrum from Ref.\ \cite{Armenise:79a}, and normalized to the GGM~1979
         data.
         The energy range and estimated mean energy are given in the legends.
         Shaded band represents $1\sigma$ variation from the average due to
         uncertainties in $M_A$ and normalization factor $N$.
        }
\label{Fig:dNQES_dQ2_Armenise_GGM79_1}
\vspace*{5mm}
\centering
\includegraphics[width=0.51\linewidth]{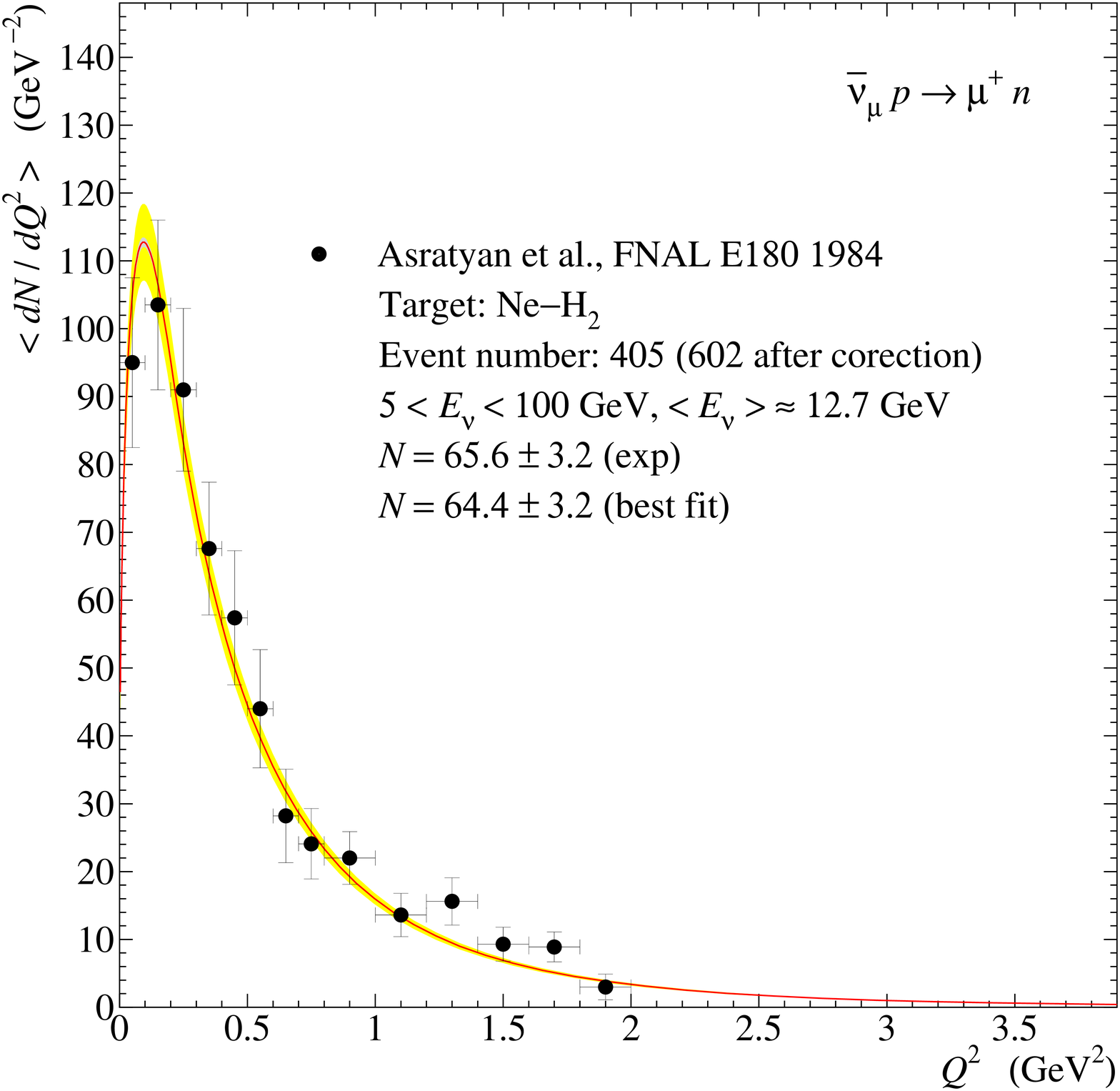}
\caption{Flux-weighted $Q^2$ distribution for $\overline\nu_{\mu}p\to\mu^+n$
         measured in the FNAL E180 experiment with a 15' bubble chamber filled
         with heavy neon--hydrogen mixture (64\% of neon atoms) and exposed to the
         FNAL wide-band $\overline\nu_{\mu}$ beam \cite{Asratyan:84a,Asratyan:84b}
         (see also Ref.\ \cite{Asratyan:82} for an earlier version).
         The curve is the distribution calculated at the mean antineutrino energy of
         $12.7\pm0.2$~GeV, with $M_A$ obtained from the global fit and then normalized
         to the E180 data. [The spectrum averaging procedure cannot be applied here,
         since the $\overline\nu_{\mu}$ spectrum has been evaluated just from the
         quoted $Q^2$ distribution.]
         Shaded band represents $1\sigma$ variation from the average due to
         uncertainties in $M_A$ and normalization factor $N$.
        }
\label{Fig:dNQES_dQ2_Asratyan_FNAL84_1}
\end{figure*}

We use the measured $Q^2$ distributions for a consistency test of our analysis.
For illustration, we show the four sets of data on $Q^2$ distributions measured in experiments
HLBC 1969 (propane) \cite{Budagov:69c} (Fig.\ \ref{Fig:dNQES_dQ2_Budagov_HLBC69_1}),
IHEP SKAT 1981 (freon) \cite{Makeev:81} (Fig.\ \ref{Fig:dNQES_dQ2_Makeev_SKAT81_1}),
CERN GGM 1979 (propane--freon mixture) \cite{Armenise:79a} (Fig.\ \ref{Fig:dNQES_dQ2_Armenise_GGM79_1}), and
FNAL E180 (neon--hydrogen mixture) \cite{Asratyan:84a,Asratyan:84b} (Fig.\ \ref{Fig:dNQES_dQ2_Asratyan_FNAL84_1}).
The curves shown in the figures are calculated with the global-fit $M_A$ and normalized to the data after
fitting of the normalization factor $N$. The shaded bands indicate the uncertainty due mainly to
indetermination of this factor.
The obtained best-fit values of $N$ should be compared with these evaluated directly from the experimental
data (all values are shown in the legends of the figures). One can see that the agreement is excellent everywhere.
So, we may conclude that this test was quite successful.

Another important confirmation of our result is a reasonably good agreement with the $M_A$ value extracted
in our earlier analysis of the data on total inelastic $\nu_{\mu}N$ and $\overline\nu_{\mu}N$ CC cross sections
and relevant observables \cite{Kuzmin:06}.

Finally, Fig.\ \ref{Fig:sQES_emt_101.0.00.301.01_1_BBBA25} presents a comparison of the total QES cross
sections for $\nu_e$, $\nu_\mu$, $\nu_\tau$, $\overline\nu_e$, $\overline\nu_\mu$, and $\overline\nu_\tau$
interactions with free nucleons, calculated with the obtained best-fit value of $M_A=0.999\pm0.011~\text{GeV}$
by using the BBBA(07) model of vector form factors. The shaded bands reproduce the uncertainty due to the $1\sigma$
error in $M_A$. 

\section{Discussion and conclusions}

We performed a statistical study of the QES total and differential cross section data in order
to extract the best-fit values of the parameters $M_A$.
Our main results are summarized in Table \ref{table:M_A} are, of course, model dependent and can
be recommended for use only within the same (or numerically equivalent) model assumptions as
in the present analysis. The best-fit values of the axial mass obtained by different fits
do not contradict to each other and agree with the recent re-extraction of $M_A$ from
$\nu_{\mu}d$, $\overline\nu_{\mu}\text{H}$, and pion electroproduction experiments, reported
in Ref.\ \cite{Bodek:07c}. They are also in agreement with the \emph{preliminary} result
of high-statistical NOMAD experiment at CERN, as well as with the numerous earlier data
which were not included into the likelihood analysis.
It has been demonstrated that removing the data subsets obtained in experiments with non-active
targets, particularly the NuTeV dataset, leads to a further \emph{decrease} of the extracted
values of $M_A$ (see Table \ref{table:M_A_Bodek}). In other words, there is no way to increase
the $M_A$ value which follows from essentially all (anti)neutrino data on total and differential
QES cross sections.

On the other hand, our best-fit value of $M_A$ is in a conflict with the mean values of $M_A$
reported by K2K and MiniBooNE Collaborations \cite{Gran:06,Aguilar-Arevalo:07},
even after accounting for the maximum possible systematic error of our analysis
related primarily to its susceptibility to the choice of the data subsets.
To expound the problem, let us consider the representative K2K result with more details.

The $M_A$ value reported in Ref.\ \cite{Gran:06} has been obtained with a water target by fitting
the $Q^2$ distributions of muon tracks reconstructed from neutrino-oxygen quasielastic interactions
by using the combined K2K-I and K2K-IIa data from the Scintillating Fiber detector (SciFi) in the
KEK accelerator to Kamioka muon neutrino beam.
The experimental data from the continuation of the K2K-II period were not used in the analysis of
Ref.\ \cite{Gran:06}.
The best-fit values of $M_A$ obtained from the K2K-I and K2K-IIa data subsets separately are,
respectively,
$1.12\pm0.12~\text{GeV}$ ($\chi^2/\text{NDF}=150/127$)
and
$1.25\pm0.18~\text{GeV}$ ($\chi^2/\text{NDF}=109/101$).

\begin{figure*}[t]
\centering
\includegraphics[width=\linewidth]{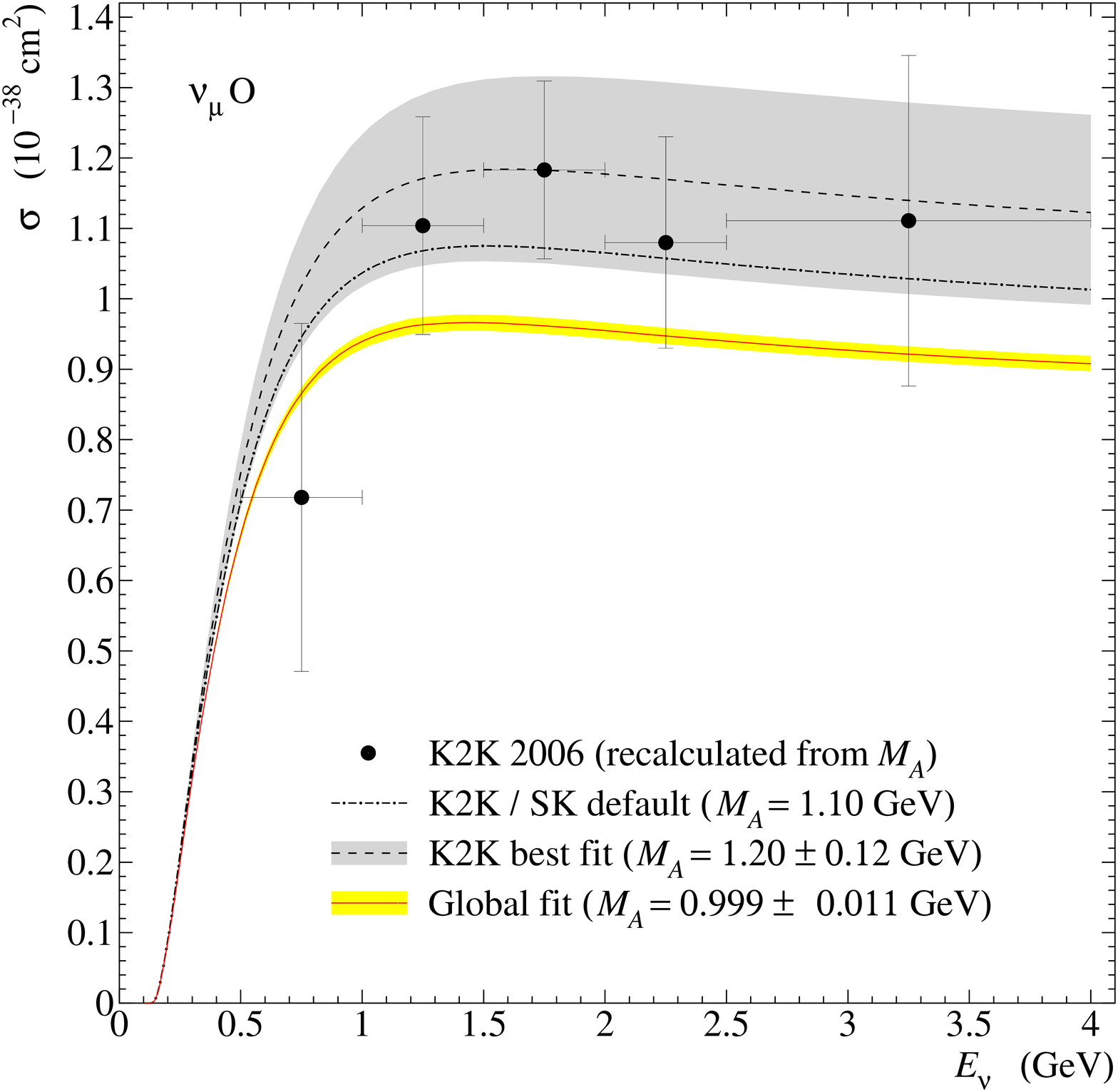}
\caption{Comparison between the QES $\nu_{\mu}$ cross sections per neutron bound in oxygen,
         evaluated with several values of the axial mass.
         The solid curve with narrow band is calculated with our best fit value of $M_A$;
         the dashed curve with wide band corresponds to the K2K extraction of $M_A$ \cite{Gran:06};
         the dash-dotted curve is calculated with the current K2K and Super-Kamiokande\,I default
         $M_A=1.1~\text{GeV}$ \cite{Ahn:06,Ashie:05}.
         The points represent the K2K cross section reconstructed (with our version of RFG model
         and BBBA(07) vector form factors) from the best-fit values of $M_A$ extracted for the
         five energy bins, as quoted in Fig.~9 of Ref.\ \cite{Gran:06}.
        }
\label{Fig:sQES_K2K06_101.0.00.301.01_1_BBBA25_SM_PRD}
\end{figure*}

Figure \ref{Fig:sQES_K2K06_101.0.00.301.01_1_BBBA25_SM_PRD} shows the $\nu_{\mu}n\to\mu^-p$
total cross section per neutron bound in oxygen, recalculated from the fitted values of $M_A$
derived in Ref.~\cite{Gran:06} from the $Q^2$ distribution shape for each reconstructed neutrino
energy.
It is necessary to underline here that the authors do not consider their result for each energy
bin as a \emph{measurement}, but rather a \emph{consistency test}.
All calculations represented in Fig.\ \ref{Fig:sQES_K2K06_101.0.00.301.01_1_BBBA25_SM_PRD}
were done with our default inputs that introduces an uncertainty of at most $2\%$;
this uncertainty is added quadratically to the quoted error bars.      
Also shown are the cross sections evaluated by using our best fit value \eqref{M_A_Global},
the K2K value of $1.20\pm0.12~\text{GeV}$, and the value of $1.1$~GeV used as a default
in the recent neutrino oscillation analyses to the data from K2K \cite{Ahn:04,Ahn:06} and
Super-Kamiokande~I \cite{Ashie:05}. A significant systematic discrepancy is clearly seen
at $E_\nu>1~\text{GeV}$.
Since the energy region covered by the K2K analysis extends to about 4~GeV, it seems
problematic to explain this discrepancy by the inapplicability of the RFG model alone.

Considering that the low-energy K2K and MiniBooNE data are in agreement with each other
and do not contradict to the high-energy NuTeV results,
we may conclude that the new generation experiments for studying the quasielastic neutrino
and antineutrino interactions with nucleons and nuclei are of urgent necessity,
in order to resolve the inconsistencies between the old and new measurements of the
axial-vector mass.

\section{Acknowledgements}

This study is currently supported by the Russian Foundation for Basic Research under Grant
No.\ 07-02-00215-a.
The authors would like to thank Krzysztof M.\ Graczyk, Sergey A.\ Kulagin, Dmitry V.\ Naumov,
Jan T.\ Sobczyk, and Oleg V.\ Teryaev for helpful discussions.
We thank the NOMAD Collaboration for permission to use their data prior to publication
and Antonio Bueno for explaining us some points of LAr TPC experiment.
We are especially grateful to Arie Bodek for his constructive comments and suggestions.
V.\,V.\,L. is very thankful to LPNHE (Paris) for warm hospitality and financial support
during a stage of this work.


\subsection*{Note on the recent SciBar result}

In the recent paper by Espinal and S\'anchez~\cite{Espinal:07},%
\footnote{Which became available to us when this paper has already
          been published}
the nucleon axial-vector form factor has been determined from an analysis of
neutrino-Carbon interactions in the K2K detector fully active SciBar tracking
calorimeter.
The best fit value of the axial-vector mass obtained in this analysis from
the $Q^2$ distribution of events and using the BBA(04) vector form
factors~\cite{Budd:04} is
\[
M_A^{(\text{SciBar})}
=1.144\pm0.077\,(\text{fit})_{-0.072}^{+0.078}\,(\text{syst})~\text{GeV}
\]
(with $\chi^2/\text{NDF}=17.2/9$, 8/9, and 9.8/9 for, respectively, 1-track,
2-track QE, and 2-track non-QE events).
It is in agreement with $M_A$ previously measured at SciFi detector in
K2K~\cite{Gran:06}. Formally (within the quoted errors) the SciBar result does
not contradict to the world average value \eqref{M_A_Global} but does not
support it.


\clearpage

\appendix

\vfill

\vfill

\vfill

\begin{thebibliography}{9999}

  \bibitem{Liesenfeld:99}
  A.~Liesenfeld et~al.,
  \emph{``A measurement of the axial form factor of the nucleon by the $p(e,e'\pi^+)n$
  reaction at $W=1125$~MeV,''}
  Phys.\ Lett.\ B {\bf 468}, 20 (1999)
  [arXiv:nucl-ex/9911003]

  \bibitem{Bernard:01}
  V.~Bernard, L.~Elouadrhiri, Ulf-G.~Mei{\ss}ner,
  \emph{``Axial structure of the nucleon: Topical review,''}
  J.\ Phys.\ G {\bf 28}, R1 (2002)
  [arXiv:hep-ph/0107088]

  \bibitem{Bodek:07c}
  A.~Bodek, S.~Avvakumov, R.~Bradford, H.~Budd,
  \emph{``Extraction of the axial nucleon form factor from neutrino experiments
  on deuterium,''}
  arXiv:0709.3538 [hep-ex]

  \bibitem{Gran:06}                         
  R.~Gran et~al. (K2K Collaboration),
  \emph{``Measurement of the quasi-elastic axial vector mass in neutrino oxygen
  interactions,''}
  Phys.\ Rev.\ D {\bf 74}, 052002 (2006)
  [arXiv:hep-ex/0603034]

  \bibitem{Aguilar-Arevalo:07}              
  A.~A.~Aguilar-Arevalo et~al. (MiniBooNE Collaboration),
  \emph{``Measurement of muon neutrino quasi-elastic scattering on carbon,''}
  Phys.\ Rev.\ Lett.\ {\bf 100}, 032301 (2008)
  [arXiv:0706.0926 [hep-ex]]

   \bibitem{Katori:07}                      
  T.~Katori (for the MiniBooNE Collaboration),
  \emph{``Cha\-r\-ged-current interaction measurements in Mini-BooNE,''}
  AIP Conf.\ Proc.\ {\bf 967}, 123 (2007)
  [arXiv:0709.4498 [hep-ex]]
 
  \bibitem{Bosted:95}
  P.~E.~Bosted,
  \emph{``An Empirical fit to the nucleon electromagnetic form-factors,''}
  Phys.\ Rev.\ C {\bf 51}, 409 (1995)
  
  \bibitem{Budd:03}
  H.~Budd, A.~Bodek, J.~Arrington,
  \emph{``Modeling quasi-elastic form factors for electron and neutrino
  scattering,''}
  arXiv:hep-ex/0308005

  \bibitem{Bradford:06}
  R.~Bradford, A.~Bodek, H.~Budd, J.~Arrington,
  \emph{``A new parameterization of the nucleon elastic form factors,''}
  Nucl.\ Phys.\ B Proc.\ Suppl.\ {\bf 159}, 127 (2006)
  [arXiv:hep-ex/0602017]
 
  \bibitem{Smith:72}
  R.~A.~Smith, E.~J.~Moniz,
  \emph{``Neutrino reactions on nuclear targets,''}
  Nucl.\ Phys.\ B {\bf 43}, 605 (1972);
  erratum -- \emph{ibid}.\ {\bf 101}, 547 (1975)



  \bibitem{Novey:67}                        
  T.~B.~Novey,
  \emph{``II. Quasi-elastic neutrino interactions,''}
  Proc.\ Roy.\ Soc.\ Lond.\ A {\bf 301}, 113 (1967)

  \bibitem{Kustom:69a}                      
  R.~L.~Kustom et~al.,
  \emph{``Quasielastic neutrino scattering,''}
  Phys.\ Rev.\ Lett.\ {\bf 22}, 1014 (1969)
 
  \bibitem{Mann:72}                         
  W.~A.~Mann et~al.,
  \emph{``Study of the reaction $\nu+n\to\mu^-+p$,''}
  in: Proceedings of the 16th International Conference on High Energy
  Physics, National Accelerator Laboratory, Chicago-Batavia, Illinois,
  September 6--13, 1972,
  ed.\ by J.~D.~Jackson, A.~Roberts
  (National Accelerator Laboratory, Batavia, Illinois, 1973), paper \#784

  \bibitem{Mann:73}                         
  W.~A.~Mann et~al.,
  \emph{``Study of the reaction $\nu+n\to\mu^-+p$,''}
  Phys.\ Rev.\ Lett.\ {\bf 31}, 844 (1973)

  \bibitem{S.Barish:75}                     
  S.~J.~Barish et~al. (ANL-Purdue Collaboration),
  \emph{``An inclusive look at ${\nu}p$ and ${\nu}n$ charged-current reactions
   below 6~GeV,''}
  preprints COO-1428-428, ANL-HEP-CP-75-38 (unpublished)

  \bibitem{Singer:77}                       
  R.~A.~Singer,
  \emph{``Study of the reaction $\nu+n\to\mu^-+p$,''}
  in: Proceedings of the International Conference on Neutrino Physics and
  Astrophysics, ``Neutrino'77,'' Baksan Valley, USSR, June 18--24, 1977,
  ed.\ by M.~A.~Markov, G.~V.~Domogatsky, A.~A.~Komar, A.~N.~Tavkhelidze
  (Publishing office ``Nauka,'' Moscow, USSR, 1978), Vol.~{\bf 2}, p.~95

  \bibitem{S.Barish:77a}                    
  S.~J.~Barish et~al.,
  \emph{``Study of neutrino interactions in hydrogen and deuterium: description of
  the experiment and study of the reaction $\nu+d\to\mu^-+p+p_s$,''}
  Phys.\ Rev.\ D {\bf 16}, 3103 (1977)

  \bibitem{Miller:82}                       
  K.~L.~Miller et~al.,
  \emph{``Study of the reaction $\nu+d\to\mu^-+p+p_s$,''}
  Phys.\ Rev.\ D {\bf 26}, 537 (1982)


  \bibitem{Cazzoli:76}                      
  E.~G.~Cazzoli et~al.,
  \emph{``Quasi-elastic neutrino scattering and the axial vector form factor,''}
  in: Proceedings of the International Neutrino Conference, Aachen,
  West Germany, June 8--12 1976,
  ed.\ by H.~Faissner, H.~Reithler, P.~Zerwas
  (Vieweg, 1977), p.~405;
  see also preprint BNL-21677, NG-349, Brookhaven National Laboratory, 1976
  (unpublished)

  \bibitem{Cnops:78a}                       
  A.~M.~Cnops et~al.,
  \emph{``Neutrino-deuterium reactions in the 7-foot bubble chamber,''}
  in: Proceedings of the Topical Conference on Neutrino Physics at
  Accelerators, Oxford, England, July 4--7, 1978,
  ed.\ by A.~G.~Michette, P.~B.~Renton
  (Rutherford Lab, 1978), p.~62;
  see also preprint BNL-24848, OG431, Brookhaven National Laboratory, 1978
  (unpublished)

  \bibitem{Fanourakis:80}                   
  G.~Fanourakis et~al.,
  \emph{``Study of low-energy antineutrino interactions on protons,''}
  Phys.\ Rev.\ D {\bf 21}, 562 (1980)

  \bibitem{Baker:81b}                       
  N.~J.~Baker et~al.,
  \emph{``Quasielastic neutrino scattering: A measurement of the weak nucleon
  axial-vector form factor,''}
  Phys.\ Rev.\ D {\bf 23}, 2499 (1981)

  \bibitem{Kitagaki:86a}                    
  T.~Kitagaki et~al.,
  \emph{``Charged-current exclusive pion production in neutrino-deuterium
  interactions,''}
  Phys.\ Rev.\ D {\bf 34}, 2554 (1986)

  \bibitem{Kitagaki:86b}                    
  T.~Kitagaki et~al., 
  \emph{Comparison of quasielastic scattering $\nu_{\mu}n\to\mu^-p$ and
  $\Delta^{++}$ production reaction $\nu_{\mu}p\to\mu^-\Delta^{++}$ in the
  BNL 7-ft deuterium bubble chamber,''}
  in: Proceedings of the International 12th Conference on Neutrino Physics
  and Astrophysics, ``Neutrino'86,'' Sendai, Japan, June 3--8, 1986,
  ed.\ by T.~Kitagaki, H.~Yuta
  (World Scientific, 1987), p.~525;
  see also preprints BNL-39020 and CONF-8606201-6, 1986 (unpublished)

  \bibitem{Abe:86}                               
  K.~Abe et~al.,
  \emph{``Precise determination of $\sin^2\theta_W$ from measurements of the
  differential cross sections for $\nu_{\mu}p\to\nu_{\mu}p$ and
  $\overline\nu_{\mu}p\to\overline\nu_{\mu}p$,''}
  Phys.\ Rev.\ Lett.\ {\bf 56}, 1107 (1986)

  \bibitem{Ahrens:87}                            
  L.~A.~Ahrens \emph{et~al.},
  \emph{``Measurement of neutrino-proton and antineutrino-proton elastic
  scattering,''}
  Phys.\ Rev.\ D {\bf 35}, 785 (1987)

  \bibitem{Ahrens:88}                       
  L.~A.~Ahrens et~al.,
  \emph{``A study of the axial-vector form factor and second-class currents in
  antineutrino quasielastic scattering,''}
  Phys.\ Lett.\ B {\bf 202}, 284 (1988)

  \bibitem{Kitagaki:90}                     
  T.~Kitagaki et~al.,
  \emph{``Study of ${\nu}d\to\mu^-pp_s$ and ${\nu}d\to\mu^-\Delta^{++}(1232)n_s$
  using the BNL 7-foot deuterium-filled bubble chamber,''}
  Phys.\ Rev.\ D {\bf 42}, 1331 (1990)

  \bibitem{Sakuda:03}                       
  M.~Sakuda,
 \emph{``Study of neutrino-nucleus interactions for neutrino oscillations,''}
  in: Proceedings of the 4th International Workshop on Neutrino Oscillations
  and Their Origin (``NOON\,2003''), Kanazawa, Japan, February 10--14, 2003,
  ed.\ by Y.~Suzuki, M.~Nakahata, M.~Shiozawa, Y.~Obayashi
  (River Edge, World Scientific, 2004), p.~253

  \bibitem{Furuno:03}                       
  K.~Furuno et~al.,
 \emph{``BNL 7-foot bubble chamber experiment -- neutrino deuterium
  interactions,''}
  a talk at the 2nd International Workshop on Neutrino-Nucleus Interactions
  in the few-GeV Region, NuInt'02,'' University of California, Irvine, December
  12--15, 2002, RCNS-03-01, KEK Preprint 2003-48, September, 2003 (unpublished)


  \bibitem{Asratyan:82}                     
  A.~E.~Asratyan et~al., 
  \emph{``Antineutrino quasielastic scattering in neon and total cross section
  for charged current interactions in the energy range 10 to 50~GeV,''}
  in: Proceedings of the 12th International Neutrino Conference
  ``Neutrino'82,'' Balatonf\"ured, Hungary, June 14--19, 1982,
  ed.\ by A.~Frenkel, L.~Jenik
  (Central Research Institute of Physics, Budapest, 1982), Supplement
  Vol.~{\bf 2}, p.~139

  \bibitem{Kitagaki:83}                     
  T.~Kitagaki et~al.,
  \emph{``High-energy quasielastic $\nu_{\mu}n\to\mu^-p$ scattering in
  deuterium,''}
  Phys.\ Rev.\ D {\bf 28}, 436 (1983)

  \bibitem{Asratyan:84a}                    
  A.~E.~Asratyan et~al.,
  \emph{``Antineutrino quasielastic scattering in neon and total cross sections
  in the energy interval 10--50~GeV,''}
  Yad.\ Fiz.\ {\bf 39}, 619 (1984)
  [Sov.\ J.\ Nucl.\ Phys.\ {\bf 39}, 392 (1984)]

  \bibitem{Asratyan:84b}                    
  A.~E.~Asratyan et~al.,
  \emph{``Total antineutrino--nucleon charged current cross section in the
   energy range 10--50~GeV,''}
  Phys.\ Lett.\ {\bf 137}\,B, 122 (1984)

  \bibitem{Ammosov:86}                      
  V.~V.~Ammosov et~al.,
  \emph{``Quasielastic production of $\Lambda$ hyperon in antineutrino
  interactions at high energies,''}
  Pisma Zh.\ Eksp.\ Teor.\ Fiz.\ {\bf 43}, 554 (1986)
  [JETP Lett.\ {\bf 43}, 716 (1986)]

  \bibitem{Ammosov:87b}                     
  V.~V.~Ammosov et~al. (IHEP-ITEP-MPEI Collaboration),
  \emph{``Neutral strange particle exclusive production in charged current
  high-energy antineutrino interactions,''}
  Z.\ Phys.\ C {\bf 36}, 377 (1987)

  \bibitem{Suwonjandee:04}                  
  N.~Suwonjandee,
  \emph{``The measurement of the quasi-elastic neutrino-nucleon scattering cross
  section at the Tevatron,''}
  Ph.\,D.\ Thesis, University of Cincinnati, Cincinnati, 2004,
  FERMILAB-THESIS-2004-67, Fermi National Accelerator Laboratory, Illinois,
  2004 (unpublished); UMI~31-20857


  \bibitem{Burmeister:65}                   
  H.~Burmeister et~al., 
  \emph{``Further analysis of the neutrino interactions in the CERN heavy liquid
  bubble chamber,''}
  in: Proceedings of the Informal Conference on Experimental Neutrino
  Physics, CERN, Geneva, January 20--22, 1965,
  ed.\ by C.~Franzinetti, CERN Yellow Report No.\ 65-32,
  European Organization for Nuclear Research, Geneva, 1965, p.~25

  \bibitem{Franzinetti:66}                  
  C.~Franzinetti,
  \emph{``Neutrino interactions in the CERN heavy liquid bubble chamber,''}
  Lecture given at the Chicago Meeting of the American Physical Society,
  Chicago, October 28, 1965, CERN Yellow Report No.\ 66-13,
  European Organization for Nuclear Research, Geneva, March 1966 (unpublished)

  \bibitem{Young:67}                        
  E.~C.~M.~Young,
  \emph{``High-energy neutrino interactions,''}
  Ph.\,D.\ Thesis,
  CERN Yellow Report No.\ 67-12,
  European Organization for Nuclear Research, Geneva, 1967 (unpublished)

  \bibitem{Orkin-Lecourtois:67}             
  A.~Orkin-Lecourtois, C.~A.~Piketty,
  \emph{``The quasi-elastic events of the CERN bubble chamber neutrino
  experiment and determination of the axial form factor,''}
  Nuovo Cim.\ {\bf 50}\,A, 927 (1967)

  \bibitem{Holder:68}                       
  M.~Holder et~al.,
  \emph{``Spark-chamber study of elastic neutrino interactions,''}
  Nuovo Cim.\ {\bf 57}\,A, 338 (1968)

  \bibitem{Budagov:69c}                     
  I.~Budagov et~al.,
  \emph{``A study of the elastic neutrino process $\nu+n\to\mu^-+p$,''}
  Lett.\ Nuovo Cim.\ {\bf 2}, 689 (1969)

  \bibitem{Eichten:72}                      
  T.~Eichten et~al.,
  \emph{``Observation of ``elastic'' hyperon production by antineutrinos,''}
  Phys.\ Lett.\ {\bf 40}\,B, 593 (1972)

  \bibitem{Eichten:73a}                     
  T.~Eichten et~al.,
  \emph{``Measurement of the neutrino--nucleon and antineutrino--nucleon total
  cross sections,''}
  Phys.\ Lett.\ {\bf 46}\,B, 274 (1973)

  \bibitem{Sciulli:74a}                     
  F.~J.~Sciulli, 
  \emph{``Total and differential cross-sections in deep inelastic neutrino
  scattering,''}
  in: Proceedings of the 4th International Conference on Neutrino Physics
  and Astrophysics ``Neutrino'74,'' Downigtown, Pennsylvania, April 26--28,
  1974,
  ed.\ by Ch.~Baltay,
  AIP Conf.\ Proc.\ {\bf 22}, 166 (1974)

  \bibitem{Haguenauer:74}                   
  M.~Haguenauer
  (for the Aachen-Brussels-CERN-Paris-Milano-Orsay-London Collaboration),
  \emph{``'Gargamelle' experiment,''}
  in: Proceedings of the 17th International Conference on High Energy
  Physics, London, July 1--10, 1974,
  ed.\ by J.~R.~Smith
  (Rutherford High Energy Laboratory, Didcot, Berkshire, England, 1975),
  p.~IV-95

  \bibitem{Rollier:75}                      
  M.~Rollier (for the Aachen-Bruxelles-CERN-Ecole Po\-ly\-technique-Orsay-London
   Collaboration),
  \emph{``Elastic neutrino and antineutrino interactions,''}
  in: Proceedings of the International Colloquium on High Energy Neutrino
  Physics, Paris, France, March 18--20, 1975,
  (Editions du CNRS, \'Ecole Polytechnique, 1975), p.~349

  \bibitem{Deden:75}                        
  H.~Deden et~al. (Gargamelle Neutrino Collaboration),
  \emph{``Experimental study of structure functions and sum rules in
  charge-changing interactions of neutrinos and antineutrinos on nucleons,''}
  Nucl.\ Phys.\ B {\bf 85}, 269 (1975)

  \bibitem{Bonetti:77}                      
  S.~Bonetti et~al.,
  \emph{``Study of quasielastic reactions of neutrino and antineutrino
  in Gargamelle,''}
  Nuovo Cim.\ {\bf 38}\,A, 260 (1977)

  \bibitem{Erriquez:77}                     
  O.~Erriquez et~al.,
  \emph{``Strange particle production by antineutrinos,''}
  Phys.\ Lett.\ {\bf 70}\,B, 383 (1977)

  \bibitem{Rollier:78}                      
  M.~Rollier (for the Gargamelle Antineutrino Collaboration,
  Bari-Milano-Strasbourg-Torino-University College London),
  \emph{``Recent results from the Gargamelle $\overline{\nu}$ propane experiment
  at the CERN-PS,''}
  in: Proceedings of the Topical Conference on Neutrino Physics at
  Accelerators, Oxford, England, July 4--7, 1978,
  ed.\ by A.~G.~Michette, P.~B.~Renton
  (Rutherford Lab, 1978), p.~68

  \bibitem{Dewit:78}                       
  M.~Dewit (for the Aachen-Bruxelles-CERN-Ecole Po\-ly\-technique-Orsay-Padova
  Collaboration),
  \emph{``Experimental study of the reaction ${\nu}n\to\mu^-p$,''}
  in: Proceedings of the Topical Conference on Neutrino Physics at
  Accelerators, Oxford, England, July 4--7, 1978,
  ed.\ by A.~G.~Michette, P.~B.~Renton
  (Rutherford Lab, 1978), p.~75

  \bibitem{Erriquez:78a}                    
  O.~Erriquez et~al.,
  \emph{``Production of strange particles in antineutrino interactions
  at the CERN PS,''}
  Nucl.\ Phys.\ B {\bf 140}, 123 (1978)

  \bibitem{Pohl:79b}                        
  M.~Pohl et~al. (Gargamelle Neutrino Propane Collaboration),
  \emph{``Experimental study of the reaction ${\nu}n\to\mu^-p$,''}
  Lett.\ Nuovo Cim.\ {\bf 26}, 332 (1979)

  \bibitem{Armenise:79a}                    
  N.~Armenise et~al.,
  \emph{``Charged current elastic antineutrino interactions in propane,''}
  Nucl.\ Phys.\ B {\bf 152}, 365 (1979)

  \bibitem{Allasia:90}                      
  D.~Allasia et~al. (Amsterdam-Bergen-Bologna-Padova-Pisa-Saclay-Torino
  Collaboration),
  \emph{``Investigation of exclusive channels in $\nu/\overline{\nu}$-deuteron
  charged current interactions,''}
  Nucl.\ Phys.\ B {\bf 343}, 285 (1990)

  \bibitem{Petti:04}                        
  R.~Petti (for the NOMAD Collaboration),
  \emph{``Precision measurements from the NOMAD experiment,''}
  in: Proceedings of 32nd International Conference on High-Energy Physics
  (ICHEP'04), Beijing, China, August 16-22, 2004,
  ed.\ by H.~Chen, D.~Du, W.~Li, C.~Lu
  (Hackensack, World Scientific, 2005), Vol.~{\bf 1}, p.~468
  [arXiv:hep-ex/0411032]

  \bibitem{Lyubushkin:06}                   
  V.~V.~Lyubushkin, B.~A.~Popov,
  \emph{``A study of quasielastic neutrino interactions $\nu_{\mu}n\to\mu^-p$
  in the NOMAD experiment,''}
  Yad.\ Fiz.\ {\bf 69}, 1917 (2006)
  [Phys.\ Atom.\ Nucl.\ {\bf 69}, 1876 (2006)]

  \bibitem{Martinez:07}                     
  A.~Martinez de la Ossa Romero,
  \emph{``Study of accelerator neutrino interactions in a liquid argon TPC,''}
  arXiv:hep-ex/0703026

  \bibitem{Lyubushkin:08a}                 
  V.~V.~Lyubushkin,
  Ph.\,D.\ Thesis, JINR, Dubna, 2008 (in preparation)

  

  \bibitem{Makeev:81}                       
  V.~V.~Makeev et~al.,
  \emph{``Quasielastic neutrino scattering $\nu_{\mu}n\to\mu^-p$ at 2 to 20~GeV
  in bubble chamber SKAT,''}
  Pisma Zh.\ Eksp.\ Teor.\ Fiz.\ {\bf 34}, 418 (1981)
  [JETP Lett.\ {\bf 34}, 397 (1981)]

  \bibitem{Belikov:81}                      
  S.~V.~Belikov et~al. (IHEP-ITEP Collaboration), 
  \emph{``Quasielastic neutrino and antineutrino interactions at the Serpukhov
  accelerator,''}
  preprint IFVE~81-146 ONF SERP-E-45, Serpukhov, 1981 (unpublished)

  \bibitem{Belikov:82a}                     
  S.~V.~Belikov et~al., 
  \emph{``Neutrino and antineutrino quasielastic scattering at 3 to 30~GeV,''}
  preprint IFVE~82-107 ONF SERP-E-45, Serpukhov, 1982 (unpublished)

  \bibitem{Belikov:82b}                     
  S.~V.~Belikov et~al.,
  \emph{``Quasielastic $\nu_{\mu}n$ scattering at energy 3--30~GeV,''}
  Yad.\ Fiz.\ {\bf 35}, 59 (1982)
  [Sov.\ J.\ Nucl.\ Phys.\ {\bf 35}, 35 (1982)]

  \bibitem{Belikov:83}                      
  S.~V.~Belikov et~al.,
  \emph{``Bounds on neutrino oscillation parameters from quasielastic scattering
  in the Serpukhov neutrino beams,''}
  Pisma Zh.\ Eksp.\ Teor.\ Fiz.\ {\bf 38}, 547 (1983)
  [JETP Lett.\ {\bf 38}, 661 (1983)]

  \bibitem{Belikov:85a}                     
  S.~V.~Belikov et~al.,
  \emph{``Restraints on parameters of oscillations of muon neutrinos from
  quasielastic scattering data,''}
  Yad.\ Fiz.\ {\bf 41}, 919 (1985)
  [Sov.\ J.\ Nucl.\ Phys.\ {\bf 41}, 589 (1985)]

  \bibitem{Belikov:85b}                     
  S.~V.~Belikov et~al.,
  \emph{``Quasielastic neutrino and antineutrino scattering:
  total cross-sections, axial-vector form-factor,''}
  Z.\ Phys.\ A {\bf 320}, 625 (1985)

  \bibitem{Grabosch:86b}                    
  H.~J.~Grabosch et~al. (SKAT Collaboration),
  preprints PHE~86-11, Berlin-Zeuthen, 1986 and
  IFVE~86-221 ONF SERP-E-107, Serpukhov, 1986 (unpublished)

  \bibitem{Grabosch:88}                     
  H.~J.~Grabosch et~al.,
  \emph{``Study of the quasielastic reactions ${\nu}n\to\mu^-p$ and
  $\overline{\nu}p\to\mu^+n$ in the SKAT bubble chamber at energies 3--20~GeV,''}
  Yad.\ Fiz.\ {\bf 47}, 1630 (1988)
  [Sov.\ J.\ Nucl.\ Phys.\ {\bf 47}, 1032 (1988)]

  \bibitem{Brunner:90}                      
  J.~Brunner et~al. (SKAT Collaboration),
  \emph{``Quasielastic nucleon and hyperon production by neutrinos and
  antineutrinos with energies below 30~GeV,''}
  Z.\ Phys.\ C {\bf 45}, 551 (1990)

  \bibitem{Bodek:07ab}
  A.~Bodek, S.~Avvakumov, R.~Bradford, H.~Budd,
  \emph{``Duality constrained parameterization of vector and axial nucleon
  form factors,''}
  Eur.\ Phys.\ J.\ C {\bf 53}, 349 (2008)
  [arXiv:0708.1946 [hep-ex]];
  see also
  A.~Bodek, S.~Avvakumov, R.~Bradford, H.~Budd,
  \emph{``Modeling atmospheric neutrino interactions: Duality constrained
  parameterization of vector and axial nucleon form factors,''}
  arXiv:0708.1827 [hep-ex]

  \bibitem{Lomon:06}
  E.~L.~Lomon,
  \emph{``Effect of revised $R_n$ measurements on extended Gari-Kr\"uempelmann
  model fits to nucleon electromagnetic form factors,''}
  arXiv:nucl-th/0609020

  \bibitem{Strumia:03}
  A.~Strumia, F.~Vissani,
  \emph{``Precise quasielastic neutrino/nucleon cross-section,''}
  Phys.\ Lett.\ B {\bf 564}, 42 (2003)
  [arXiv:astro-ph/0302055]

  \bibitem{Kuzmin:05b}
  K.~S.~Kuzmin, V.~V.~Lyubushkin, V.~A.~Naumov,
  \emph{``Tau lepton polarization in quasielastic neutrino-nucleon scattering,''}
  Nucl.\ Phys.\ B Proc.\ Suppl.\ {\bf 139}, 154 (2005)
  [arXiv:hep-ph/0408107]

  \bibitem{Kuzmin:06b}
  K.~S.~Kuzmin, V.~V.~Lyubushkin, V.~A.~Naumov,
  \emph{``Axial masses in quasielastic neutrino scattering and single-pion
  neutrinoproduction on nucleons and nuclei,''}
  Acta Phys.\ Polon.\ B {\bf 37}, 2337 (2006)
  [arXiv:hep-ph/0606184]

  \bibitem{LlewellynSmith:72}
  C.~H.~Llewellyn~Smith,
  \emph{``Neutrino reactions at accelerator energies,''}
  Phys.\ Rept.\ {\bf 3}\,C, 261 (1972)

  \bibitem{Wilkinson:00-01}
  D.~H.~Wilkinson,
  \emph{``Limits to second-class nucleonic and mesonic currents,''}
  Eur.\ Phys.\ J.\ A {\bf 7}, 307 (2000);
  D.~H.~Wilkinson,
  \emph{``Limits to second-class nucleonic currents,''}
  Nucl.\ Instrum.\ Meth.\ A {\bf 456}, 655 (2000);
  D.~H.~Wilkinson,
  \emph{``Second-class currents and $\Delta{s}$ in $\nu(\overline{\nu})p$
  elastic scattering,''}
  \emph{ibid}., A {\bf 469}, 286 (2001)

  \bibitem{Gardner:01}
  S.~Gardner, C.~Zhang,
  \emph{``Sharpening low-energy, standard-model tests via correlation
  coefficients in neutron $\beta$ decay,''}
  Phys.\ Rev.\ Lett.\ {\bf 86}, 5666 (2001)
  [arXiv:hep-ph/0012098]

  \bibitem{Kelly:04}
  J.~J.~Kelly,
  \emph{``Simple parametrization of nucleon form factors,''}
  Phys.\ Rev.\  C {\bf 70}, 068202 (2004)

  \bibitem{Gari:92}
  M.~F.~Gari, W.~Kr\"uempelmann,
  \emph{``The electric neutron form-factor and the strange quark content of
  the nucleon,''}
  Phys.\ Lett.\ B {\bf 274}, 159 (1992);
  \emph{ibid.} {\bf 282}, 483 (E) (1992)

  \bibitem{Lomon:01}
  E.~L.~Lomon,
  \emph{``Extended Gari-Kr\"uempelmann model fits to nucleon electromagnetic
  form factors,''}
  Phys.\ Rev.\ C {\bf 64}, 035204 (2001)
  [arXiv:nucl-th/0104039]

  \bibitem{Lomon:02}
  E.~L.~Lomon,
  \emph{``Effect of recent $R_p$ and $R_n$ measurements on extended
  Gari-Kr\"uempelmann model fits to nucleon electromagnetic form factors,''}
  Phys.\ Rev.\ C {\bf 66}, 045501 (2002)
  [arXiv:nucl-th/0203081]

  \bibitem{Arrington:03a}
  J.~Arrington,
  \emph{``How well do we know the electromagnetic form factors of the proton?,''}
  Phys.\ Rev.\ C {\bf 68}, 034325 (2003)
  [arXiv:nucl-ex/0305009]

  \bibitem{Arrington:03b}
  J.~Arrington,
  \emph{``How well do we know the electromagnetic form factors of the proton?,''}
  Eur.\ Phys.\ J.\ A {\bf 17}, 311 (2003)
  [arXiv:hep-ph/0209243]

  \bibitem{Finn:04}
  J.~M.~Finn (for the JLab E93-038 Collaboration),
  \emph{``Measurements of the electric form factor of the neutron at JLab via
  recoil polarimetry in the reaction: $d(\vec{e},e'\vec{n})p$,''}
  Fizika B {\bf 13}, 545 (2004)

  \bibitem{Anderson:07}
  B.~Anderson~\emph{et~al.} (Jefferson Lab E95-001 Collaboration),
  \emph{``Extraction of the neutron magnetic form factor from quasi-elastic
  ${}^3\vec{He}(\vec{e},e')$ $Q^2=0.1-0.6~(\text{GeV}/c)^2$,''}
  Phys.\ Rev.\ C {\bf 75}, 034003 (2007)
  [arXiv:nucl-ex/0605006]

  \bibitem{Day:07}
  D.~Day,
  \emph{``Nucleon elastic form factors: Current status of the experimental
  effort,''}
  Eur.\ Phys.\ J.\ A {\bf 31}, 560 (2007)

  \bibitem{Perdrisat:07}
  C.~F.~Perdrisat, V.~Punjabi, M.~Vanderhaeghen,
  \emph{``Nucleon electromagnetic form factors,''}
  Prog.\ Part.\ Nucl.\ Phys.\ {\bf 59}, 694 (2007)
  [arXiv:hep-ph/0612014]


  \bibitem{Hagiwara:04}
  K.~Hagiwara, K.~Mawatari, H.~Yokoya,
  \emph{``Pseudoscalar form factors in tau-neutrino nucleon scattering,''}
  Phys.\ Lett.\ B {\bf 591}, 113 (2004)
  [arXiv:hep-ph/0403076]

  \bibitem{Kuzmin:04}
  K.~S.~Kuzmin, V.~V.~Lyubushkin, V.~A.~Naumov,
  \emph{``Polarization of tau leptons produced in quasielastic neutrino nucleon
  scattering,''}
  Mod.\ Phys.\ Lett.\ A {\bf 19}, 2919 (2004)
  [arXiv:hep-ph/0403110]

  \bibitem{Yao:06}
  W.~M.~Yao et~al. (Particle Data Group),
  \emph{``Review of particle physics,''}
  J.\ Phys.\ G {\bf 33}, 1 (2006)
  and 2007 partial update for the 2008 edition,
  see URL \url{http://pdg.lbl.gov/}

  \bibitem{Pocanic:04}
  D.~Pocanic et~al.,
  \emph{``Precise measurement of the $\pi^+\to\pi^0e^+\nu$ branching ratio,''}
  Phys.\ Rev.\ Lett.\ {\bf 93}, 181803 (2004)
  [arXiv:hep-ex/0312030]

  \bibitem{Nakamura:02}
  S.~Nakamura et~al.
  \emph{``Neutrino deuteron reactions at solar neutrino energies,''}
  Nucl.\ Phys.\  A {\bf 707}, 561 (2002)
  [arXiv:nucl-th/0201062]


  \bibitem{Perkins:72}
  D.~H.~Perkins,
  \emph{``Neutrino interactions,''}
  in: Proceedings of the 16th International Conference on High Energy
  Physics, National Accelerator Laboratory, Chicago-Batavia, Illinois,
  September 6--13, 1972,
  ed.\ by J.~D.~Jackson, A.~Roberts
  (National Accelerator Laboratory, Batavia, Illinois, 1973), Vol.~{\bf IV},
  p.~189

  \bibitem{Derrick:74}
  M.~Derrick,
  \emph{``Charged current neutrino reactions in the resonance region,''}
  in: Proceedings of the 17th International Conference on High Energy
  Physics, London, July 1--10, 1974,
  ed.\ by J.~R.~Smith
  (Rutherford High Energy Laboratory, Didcot, Berkshire, England, 1975),
  p.~II-166

  \bibitem{Perkins:75}
  D.~H.~Perkins,
  \emph{``Review of neutrino experiments,''}
  in: Proceedings of the 1975 International Symposium on Lepton and Photon
  Interactions at High Energies, Stanford University, August 21--27, 1975,
  ed.\ by T.~W.~Kirk
  (Stanford Linear Accelerator Center, Stanford, 1975), p.~571
  
  \bibitem{Cline:77}
  D.~Cline, W.~F.~Fry,
  \emph{``Neutrino scattering and new particle production,''}
  Ann.\ Rev.\ Nucl.\ Part.\ Sci.\ {\bf 27}, 209 (1977)

  \bibitem{Wachsmuth:77}
  H.~Wachsmuth,
  \emph{``Accelerator neutrino physics,''}
  Lectures held at the Herbstschule f\"ur Hochenergiephysik, Maria Laach, Eifel,
  Germany, September 14--24, 1976,
  preprint CERN/EP/PHYS~77-40, August 17, 1977 (unpublished)

  \bibitem{Ermolov:78}
  P.~F.~Ermolov, A.~I.~Mukhin,
  \emph{``Neutrino experiments at high energies,''}
  Usp.\ Fiz.\ Nauk {\bf 124}, 385 (1978)
  [Sov.\ Phys.\ Uspekhi {\bf 21}, 185-214 (1978)]

  \bibitem{Musset:78}                       
  P.~Musset, J.-P.~Vialle,
  \emph{``Neutrino physics with Gargamelle,''}
  Phys.\ Rept.\ {\bf 39}, 1 (1978)

  \bibitem{Alekhin:87}
  S.~I.~Alekhin et~al.,
  \emph{``Compilation of cross-sections IV:
  $\gamma,\nu,\Lambda,\Sigma,\Xi,\Lambda$, and $K_L^0$ induced reactions,''}
  Report CERN-HERA~87-01,
  European Organization for Nuclear Research, Geneva, 1987

  \bibitem{Sakuda:02}
  M.~Sakuda,
  \emph{``Results from low-energy neutrino nucleus scattering experiments,''}
  Nucl.\ Phys.\ B Proc.\ Suppl.\ {\bf 112}, 109 (2002)

  \bibitem{Ammosov:92}
  V.~V.~Ammosov et~al.,
  \emph{``Investigation of neutrino interactions using the bubble chamber
  SKAT,''}
  Fiz.\ Elem.\ Chast.\ Atom.\ Yadra {\bf 23}, 648 (1992)
  [Sov.\ J.\ Part.\ Nucl.\ {\bf 23}, 283 (1992)]

  \bibitem{Baltay:94}
  C.~Baltay,
  \emph{``Deep inelastic neutrino interactions and charm production,''}
  Nucl.\ Phys.\ B Proc.\ Suppl.\ {\bf 36}, 363 (1994)

  \bibitem{Zeller:03}
  G.~P.~Zeller,
  \emph{``Low energy neutrino cross sections: Comparison of various Monte Carlo
  predictions to experimental data,''}
  a talk at the 2nd International Workshop on Neutrino-Nucleus Interactions
  in the few-GeV Region, NuInt'02,'' University of California, Irvine,
  December 12--15, 2002;
  arXiv:hep-ex/0312061

  \bibitem{Fleming:06}
  B.~Fleming,
  \emph{``Neutrino cross sections and scattering physics,''}
  AIP Conf.\ Proc.\ {\bf 815}, 1 (2006)

  \bibitem{Sorel:07}
  M.~Sorel,
  \emph{``Overview of progress in neutrino scattering measurements,''}
  AIP Conf.\ Proc.\ {\bf 967}, 17 (2007)
  [arXiv:0710.3966 [hep-ex]]

  \bibitem{Gran:07}
  R.~Gran,
  \emph{``Progress in measuring neutrino quasielastic interactions,''}
  AIP Conf.\ Proc.\ {\bf 967}, 141 (2007)
  [arXiv:0711.3024 [hep-ex]]


  \bibitem{MINUIT}
  F.~James,
  \emph{``MINUIT, Reference Manual, Version 94.1,''}
  CERN Program Library Long Writeup D506
  (European Organization for Nuclear Research, Geneva, 1994);
  F.~James, M.~Roos,
  \emph{``MINUIT: A system for function minimization and analysis of the
  parameter errors and correlations,''}
  Comput.\ Phys.\ Commun.\ {\bf 10}, 343 (1975)

  \bibitem{Budagov:69a}                         
  I.~Budagov \emph{et~al.},
  \emph{``Single pion production by neutrinos on free protons,''}
  Phys.\ Lett.\ {\bf 29}\,B, 524 (1969)

  \bibitem{Allasia:84}                          
  D.~Allasia et~al.
  (Amsterdam-Bergen-Bologna-Padova-Pisa-Saclay-Torino Collaboration),
  \emph{``Measurement of the $\nu_\mu$ and $\overline{\nu}_\mu$ nucleon
  charged-current total cross sections, and the ratio of $\nu_\mu$ neutron to
  $\nu_\mu$ proton charged-current total cross sections,''}
  Nucl.\ Phys.\ B {\bf 239}, 301 (1984)

  \bibitem{Singh:72}
  S.~K.~Singh,
  \emph{``The effect of final state interactions and deuteron binding in
  ${\nu}d\to\mu^-pp$,''}
  Nucl.\ Phys.\ B {\bf 36}, 419 (1972)

  \bibitem{Baker:83a}                           
  N.~J.~Baker \emph{et~al.},
  \emph{``Exclusive neutral-current reaction $\nu_{\mu}n\to\nu_{\mu}p\pi^-$
  in the BNL 7-foot deuterium bubble chamber,''}
  Phys.\ Rev.\ D {\bf 28}, 2900 (1983)

  \bibitem{Butkevich:05}
  A.~V.~Butkevich, S.~P.~Mikheyev,
  \emph{``Test of Fermi gas model and plane-wave impulse approximation against
  electron nucleus scattering data,''}
  Phys.\ Rev.\ C {\bf 72}, 025501 (2005)
  [arXiv:hep-ph/0505008]

  \bibitem{Butkevich:07}
  A.~V.~Butkevich, S.~A.~Kulagin,
  \emph{``Quasi-elastic neutrino charged-current scattering cross sections
  on oxygen,''}
  Phys.\ Rev.\ C {\bf 76}, 045502 (2007)
  [arXiv:0705.1051 [nucl-th]];
  see also
  A.~V.~Butkevich, S.~A.~Kulagin,
  \emph{``QE neutrino CC cross sections off ${}^{16}$O,''}
  AIP Conf.\ Proc.\ {\bf 967}, 298 (2007)
  [arXiv:0711.3223 [nucl-th]]

  \bibitem{Kuzmin:06}
  K.~S.~Kuzmin, V.~V.~Lyubushkin, V.~A.~Naumov,
  \emph{``Fine-tuning parameters to describe the total charged-current
  neutrino-nucleon cross section,''}
  Yad.\ Fiz.\ {\bf 69}, 1898 (2006)
  [Phys.\ Atom.\ Nucl.\ {\bf 69}, 1857 (2006)];
  K.~S.~Kuzmin, V.~V.~Lyubushkin, V.~A.~Naumov,
  \emph{``How to sum contributions into the total charged-current
  neutrino-nucleon cross section,''}
  arXiv:hep-ph/0511308

  \bibitem{Ahn:04}
  M.~H.~Ahn et~al. (K2K Collaboration),
  \emph{``Search for electron neutrino appearance in a 250 km long baseline
  experiment,''}
  Phys.\ Rev.\ Lett.\ {\bf 93}, 051801 (2004)
  [arXiv:hep-ex/0402017]

  \bibitem{Ahn:06}
  M.~H.~Ahn et~al. (K2K Collaboration),
  \emph{``Measurement of neutrino oscillation by the K2K experiment,''}
  Phys.\ Rev.\ D {\bf 74}, 072003 (2006)
  [arXiv:hep-ex/0606032]

  \bibitem{Ashie:05}
  Y.~Ashie et~al. (Super-Kamiokande Collaboration),
  \emph{``A measurement of atmospheric neutrino oscillation para\-meters by
  Super-Kamiokande~I,''}
  Phys.\ Rev.\ D {\bf 71}, 112005 (2005)
  [arXiv:hep-ex/0501064]

\bigskip\hrule\bigskip
{\bf ADDITIONAL REFERENCES}
\bigskip\hrule\bigskip


 \bibitem{Espinal:07}
  X.~Espinal, F.~S\'anchez,
  \emph{``Measurement of the axial vector mass in neutrino-Carbon interactions
  at K2K,''}
  AIP Conf.\ Proc.\ {\bf 967}, 117 (2007)

  \bibitem{Budd:04}
  H.~Budd, A.~Bodek, J.~Arrington,
  \emph{``Vector and axial form factors applied to neutrino quasielastic
  scattering,''}
  Nucl.\ Phys.\ B Proc.\ Suppl.\ {\bf 139}, 90 (2005)
  [arXiv:hep-ex/0410055]


\end{thebibliography}
\end{document}